\documentclass[prl,preprint,showpacs,floatfix,amsmath,amssymb,nobibnotes,nolongbibliography,superscriptaddress]{revtex4-2}

\usepackage{graphicx}
\usepackage{dcolumn}
\usepackage{bm}
\usepackage{float}
\usepackage{amssymb}
\usepackage{color}
\usepackage{multirow}
\usepackage{stmaryrd}
\usepackage[colorlinks,linkcolor=blue,urlcolor=blue,citecolor=blue]{hyperref}

\setcitestyle{super}

\begin{document}


\title{Giant nonlinear optical wave mixing in van der Waals compound MnPSe$_3$}

\author{Li Yue}
\thanks{These authors contributed equally to this study.}
\email{lilyyue@pku.edu.cn}
\affiliation{International Center for Quantum Materials, School of Physics, Peking University, Beijing 100871, China}
\author{Chang Liu}
\thanks{These authors contributed equally to this study.}
\affiliation{International Center for Quantum Materials, School of Physics, Peking University, Beijing 100871, China}
\author{Shanshan Han}
\affiliation{Beijing Academy of Quantum Information Sciences, Beijing 100913, China}
\author{Hao Hong}
\affiliation{State Key Laboratory for Mesoscopic Physics, Frontiers Science Center for Nano-optoelectronics, School of Physics, Peking University, Beijing 100871, China.}
\author{Yijun Wang}
\affiliation{State Key Laboratory for Mesoscopic Physics, Frontiers Science Center for Nano-optoelectronics, School of Physics, Peking University, Beijing 100871, China.}
\author{Qiaomei Liu}
\affiliation{International Center for Quantum Materials, School of Physics, Peking University, Beijing 100871, China}
\author{Jiajie Qi}
\affiliation{State Key Laboratory for Mesoscopic Physics, Frontiers Science Center for Nano-optoelectronics, School of Physics, Peking University, Beijing 100871, China.}
\author{Yuan Li}
\affiliation{International Center for Quantum Materials, School of Physics, Peking University, Beijing 100871, China}
\affiliation{Collaborative Innovation Center of Quantum Matter, Beijing 100871, China}
\author{Dong Wu}
\affiliation{Beijing Academy of Quantum Information Sciences, Beijing 100913, China}
\author{Kaihui Liu}
\affiliation{State Key Laboratory for Mesoscopic Physics, Frontiers Science Center for Nano-optoelectronics, School of Physics, Peking University, Beijing 100871, China.}
\author{Enge Wang}
\affiliation{International Center for Quantum Materials, School of Physics, Peking University, Beijing 100871, China}
\affiliation{Collaborative Innovation Center of Quantum Matter, Beijing 100871, China}
\author{Tao Dong}
\email{taodong@pku.edu.cn}
\affiliation{International Center for Quantum Materials, School of Physics, Peking University, Beijing 100871, China}
\author{Nanlin Wang}
\email{nlwang@pku.edu.cn}
\affiliation{International Center for Quantum Materials, School of Physics, Peking University, Beijing 100871, China}
\affiliation{Beijing Academy of Quantum Information Sciences, Beijing 100913, China}
\affiliation{Collaborative Innovation Center of Quantum Matter, Beijing 100871, China}



\maketitle
\clearpage

\section*{Abstract}
Optical nonlinearities, one of the most fascinating properties of two-dimensional (2D) materials, are essential for exploring novel physics in 2D systems and developing next-generation nonlinear optical applications. While tremendous efforts have been made to discover and optimize second-order nonlinear optical responses in various 2D materials, higher odd-order nonlinear processes, which are in general much less efficient than second order ones, have been paid less attention despite their scientific and applicational significance. Here we report giant odd-order nonlinear optical wave mixing in a correlated van der Waals insulator MnPSe$_3$ at room temperature. Illuminated by two near-infrared femtosecond lasers simultaneously, it generates a series of degenerate and non-degenerate four- and six-wave mixing outputs, with conversion efficiencies up to the order of $10^{-4}$ and $10^{-6}$ for the four- and six-wave mixing processes, respectively, far exceeding the efficiencies of several prototypical nonlinear optical materials (GaSe, LiNbO$_3$).
This work highlights the intriguing prospect of transition metal phosphorous trichalcogenides for future research of the nonlinear light-matter interactions in 2D systems and for potential nonlinear photonic applications.

\clearpage
\section*{Introduction}

The coherent nonlinear interaction between light and matter gives rise to intriguing nonlinear optical wave mixing phenomena \cite{Boydbook,Shenbook}, such as high harmonic generation (HHG), sum frequency generation (SFG) and four-wave mixing (FWM). Over the past decades, nonlinear optical wave mixing plays a crucial role in laser generation and manipulation, photon detection and optical sensing \cite{Boydbook,Shenbook,Duellibook,KellerNature2003,GarmireOptExp2013,AutereAdvMater2018}. Nowadays, nonlinear optical wave mixing becomes increasingly important in emerging fields such as quantum photonics, quantum information and on-chip nanophotonics \cite{AutereAdvMater2018,ChangNatPhoto2014,ElshaariNatPhoto2020}.
As nonlinear optical responses are inherently weak, materials with large nonlinear susceptibility are essential for applications. Traditional nonlinear optical devices are typically based on conventional bulk nonlinear crystals (e.g. beta barium borate, lithium niobate). However, the conventional materials are limited by insufficiently large susceptibility and weak availability for nano-integration due to its three dimensional covalent bonding \cite{Boydbook,GarmireOptExp2013,AutereAdvMater2018}, therefore cannot satisfy the demands of next-generation nonlinear photonic applications.

The flourishing of two-dimensional (2D) materials opens up new opportunities. The van der Waals (vdW) structure of 2D materials is advantageous for nano-fabrication and bond-free integration \cite{LiuNature2019}. In recent years, large nonlinear susceptibilities have been found in a variety of 2D materials, such as graphene and transition metal dichalcogenides \cite{AutereAdvMater2018,GuoLaserPhoRev2019}. Despite tremendous research of the nonlinear optical properties of 2D materials, further exploration could be extended in two aspects.
Firstly, current research mostly focuses on the second-order nonlinear processes in non-centrosymmetic materials, whereas the inversion independent odd-order nonlinear processes are paid less attention.
For example, in some cases, third order nonlinear response is achieved via cascaded second-order processes \cite{HunaultOptLett2010,ArahiraOptExp2011,ZhangNpjQI2021}, since second-order nonlinear process is in general more effective than the third-order one. As for the one with large odd-order susceptibilities (i.e., graphene), the conversion efficiency is highly restricted by the intense absorption with increasing thickness \cite{HendryPRL2010,HongPRX2013}.
Secondly, the well-studied second harmonic generation (SHG) and HHG are degenerate wave mixing processes under monochromatic laser excitation. Non-degenerate wave mixing processes (e.g., non-degenerate four-wave mixing), which require excitation by multiple wavelengths, are seldom studied but possess unique importance. The non-degeneracy provides different excitation quantum pathways for nonlinear light-matter interactions \cite{RuedaACSPhoto2021,BauerNatPhoto2022}. The tunable time delay between different laser pulses enables tracking the temporal dynamics of the interacting system \cite{BauerNatPhoto2022}.

Transition metal phosphorous trichalcogenides MPX$_3$ (M=Mn, Ni, Fe, Co; X=S, Se) are 2D vdW correlated anti-ferromagnetic (AFM) insulators. In recent years, these compounds have attracted lots of interest due to the spin-charge-lattice correlations \cite{KimPRL2018,ErgecenNatCommu2022,WangNatMater2021,KangNature2020}, excitonic many-body quantum states \cite{KangNature2020,HwangboNatNanotech2021,BelvinNatCommu2021}, tunable magnetic and optical properties through strain \cite{NiNatNanotech2021}, layer numbers \cite{HwangboNatNanotech2021,KimNatCommu2019}, and cavities \cite{ZhangNatPhoto2022}.
So far, plenty of studies have reported their fascinating optical properties including photoluminescence \cite{WangNatMater2021,KangNature2020, HwangboNatNanotech2021}, giant linear dichroism \cite{HwangboNatNanotech2021,ZhangNanoLett2021,ZhangNatPhoto2022} and nonlinear optical second harmonics \cite{NiNatNanotech2021,NiPRL2021,ChuPRL2020,ShanNature2021,WangArxiv2023}, but all these studies focus on the AFM phases at low temperatures.
In the paramagnetic phase, even-order nonlinear responses are absent due to the inversion symmetry, but odd-order nonlinear responses should in principle exist, which, to our best knowledge, haven't been reported before.
In this work, we observed profound odd-order optical nonlinearities in centrosymmetric MnPSe$_3$ at room temperature. Excited by two femtosecond lasers at 847 nm and 1280 nm, the MnPSe$_3$ sample generates nonlinear wave mixing outputs ranging from ultraviolet to visible wavelengths, including third harmonic generation (THG), non-degenerate four-wave and six-wave mixing.


\section*{Experimental results}

\begin{figure}[htbp]
    \begin{center}
	   \includegraphics[clip, width=5.5in]{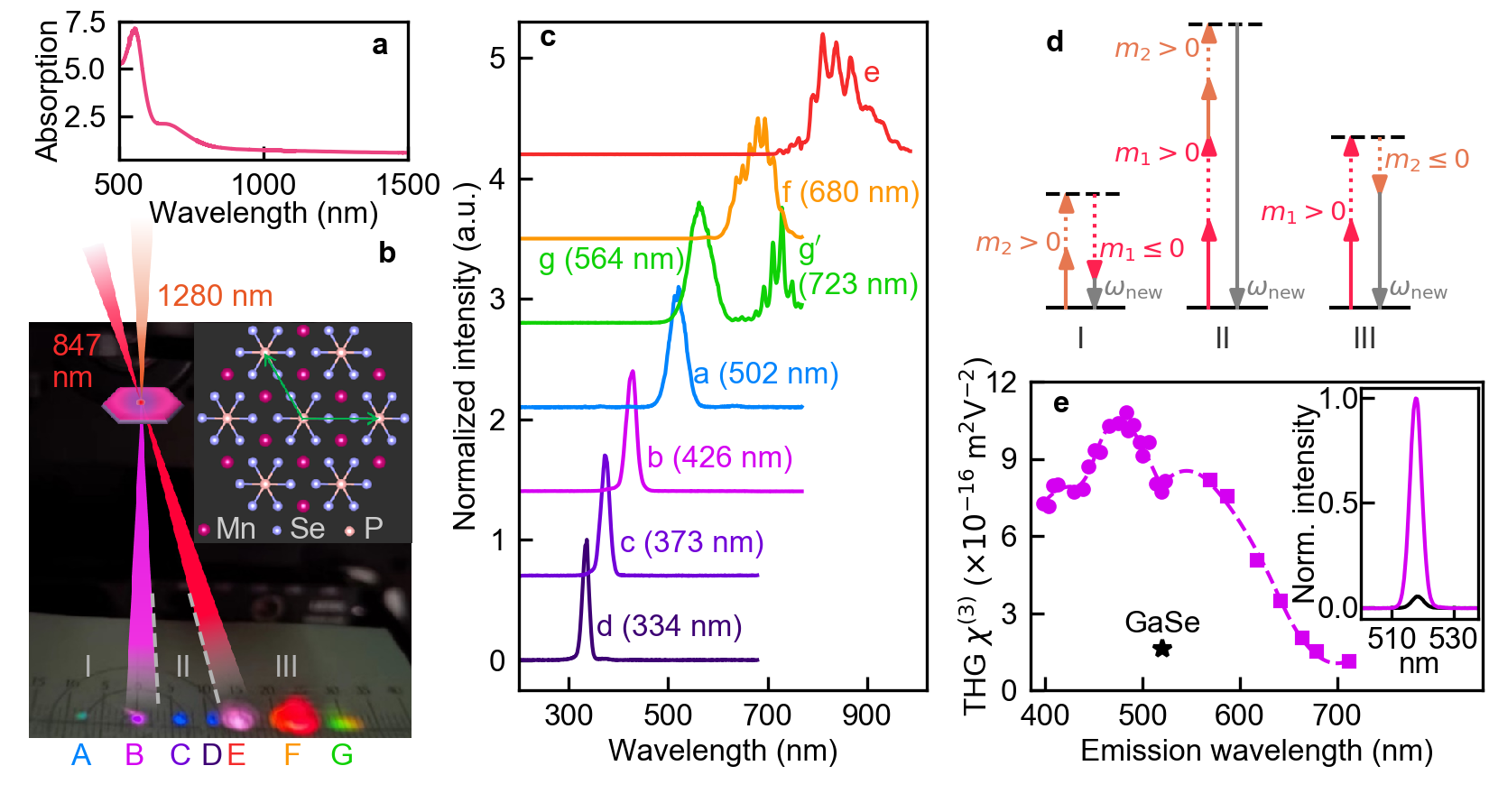} 
	   \caption{\textbf{Optical wave mixing effects in MnPSe$_3$.} \textbf{a}, Absorption spectrum of MnPSe$_3$ sample calculated as $-\mathrm{ln} (T)$, where $T$ is the measured broadband transmission ratio (see Methods). \textbf{b}, View of the nonlinear wave mixing signals of the $\sim 20$ $\mu$m MnPSe$_3$ bulk. Beam D (334 nm) and C (373 nm) are in the ultraviolet range but were seen on the paper, thanks to the fluorescent whitening agents inside the paper. The photo was taken by a cell phone camera. The inset shows the in-plane structure with green arrows representing the $a$ and $b$ axes. \textbf{c}, Normalized spectra for beam A-G, vertically shifted for clarity. The signals located at longer than 600 nm contain split sharp peaks which are likely caused by interference effect from the bulk flake sample (see Supplementary S2). \textbf{d}, Schematics of wave mixing processes of different $(m_1,m_2)$ channels, corresponding to the three regions in panel \textbf{b}. \textbf{e}, $\chi^{(3)}$ as a function of emission wavelength obtained by measuring wavelength-dependent THG spectrum (see Methods and Supplementary S4). The dashed line is a guide-to-the-eye. The circle and square markers are measured with the signal and idler output of an optical parametric amplifier. The black star denotes the $\chi^{(3)}$ value of GaSe. Inset, black and magenta lines represent the THG response measured on 80-nm thick GaSe (black line) and MnPSe$_3$ (magenta line) films, both are normalized by the peak intensity of MnPSe$_3$.}
	   \label{fig:1}
    \end{center}
\end{figure}

We fabricated several MnPSe$_3$ samples: an exfoliated bulk of $\sim 20$ $\mu$m thickness glued onto a hollow copper platform to avoid using substrates, and two thin films with thickness of $\sim1$ $\mu$m and $\sim100$ nm exfoliated on fused silica. Our experiment was performed at room temperature. Two near-infrared femtosecond lasers at 847 nm (power $P_1=5$ mW) and 1280 nm (power $P_2=10$ mW), with 50 kHz repetition rate, are focused onto a spot with $\sim 100$ $\mu$m diameter on the sample in non-collinear geometry by two lenses (see Methods and Supplementary S1). The peak intensities are approximately 0.03 TW/cm$^2$ for 847 nm laser and 0.06 TW/cm$^2$ for 1280 nm laser. The two excitation wavelengths are within the optical gap of MnPSe$_3$ and weakly absorbed (Fig.~\ref{fig:1}a). 
Fig.~\ref{fig:1}(b) demonstrates the transmitted beams through the $\sim 20$ $\mu$m bulk sample directly viewed on a paper card. When the 847 nm and 1280 nm pulses are temporally separated, only two outgoing beams are seen (speckles B and E in Fig.~\ref{fig:1}b). While E is a direct view of the 847 nm laser, the visible beam B contains the third harmonic generation (THG) of the 1280 nm laser.
When the two impinging pulses are temporally overlapped, the sample emits new outgoing beams (speckles A, C, D, F, G in Fig.~\ref{fig:1}a) arising from non-degenerate nonlinear wave mixing processes.

The nonlinear wave mixing process, $E_\mathrm{new} \propto \chi^{(m)} E_{1}^{|m_1|} E_{2}^{|m_2|}$, generates new outgoing photons satisfying energy conservation $\omega_\mathrm{new} = m_1 \omega_1+m_2 \omega_2$ (Fig.~\ref{fig:1}d). Here $E_{1}$ ($E_{2}$) and $\omega_1$ ($\omega_2$) refer to the electric field and photon energy of the 847 nm (1280 nm) laser. $\chi^{(m)}$ is the $m$-th order nonlinear susceptibility, with $m=|m_1|+|m_2|$ being odd integer due to inversion symmetry. In non-collinear geometry, new photons with momentum in the direction between the two excitation beams should be from all-sum processes (region II in Fig.~\ref{fig:1}b,d), and photons emitted in the outside region should involve a difference process (region I and III in Fig.~\ref{fig:1}b,d).
We measured the spectrum of each outgoing beam with other beams blocked using a NOVA spectrometer, ideaoptics, China (Fig.~\ref{fig:1}c). The measurements cover 300-980 nm wavelength range. The peaks are identified as all four-wave and six-wave mixing outputs within 300-980 nm as listed in Table~\uppercase\expandafter{\romannumeral1}.
Particularly, the spectrum at speckle G exhibits a second peak g$^{\prime}$ centered at 723 nm (1.72 eV), matching none of the four-wave and six-wave mixing energies.

\begin{table} [htbp]
    \begin{center}
    \begin{tabular}{|c|c|c|}
      \hline
      \multicolumn{2}{|c|}{$m=3$, four-wave mixing} \\ \hline
      $(m_1,m_2)$ channels  & corresponding peaks and measured energy (eV) \\ \hline
      $(0,3)$ &  peak b of beam B, 2.91 eV (426 nm) \\ \hline
      $(1,2)$ &  peak c of beam C, 3.2 eV (373 nm)\\ \hline
      $(2,1)$ &  peak d of beam D, 3.71 eV (334 nm)\\ \hline
      $(2,-1)$ &  peak f of beam F, 1.82 eV (680 nm)\\ \hline
      \multicolumn{2}{|c|}{others: $(-1,2)$, $(3,0)$, out of 300-980 nm range} \\
      \hline \hline
      \multicolumn{2}{|c|}{$m=5$, six-wave mixing} \\ \hline
      $(m_1,m_2)$ channels & corresponding peaks and measured energy (eV) \\ \hline
      (-1,4) &  peak a of beam A, 2.47 eV (502 nm)\\ \hline
      (3,-2) &  peak g of beam G, 2.2 eV (564 nm) \\ \hline
      \multicolumn{2}{|c|}{others: $(0,5)$, $(1,4)$, $(2,3)$, $(2,-3)$, $(3,2)$, $(4,1)$, } \\
      \multicolumn{2}{|c|}{$(4,-1)$, $(5,0)$, out of 300-980 nm range} \\
      \hline
    \end{tabular}
    \caption{\textbf{Different nonlinear wave mixing channels.} Calculated and measured energies for nonlinear wave mixing processes of third ($m=3$) and fifth order ($m=5$). $(m_1,m_2)$ values leading to $m_1 \omega_1+m_2 \omega_2<0$ are invalid and not listed in the table. See Supplementary S3 for more details.}
    \label{Table1}
\end{center}
\end{table}

\begin{figure}[htbp]
    \begin{center}
	   \includegraphics[clip, width=6.1in]{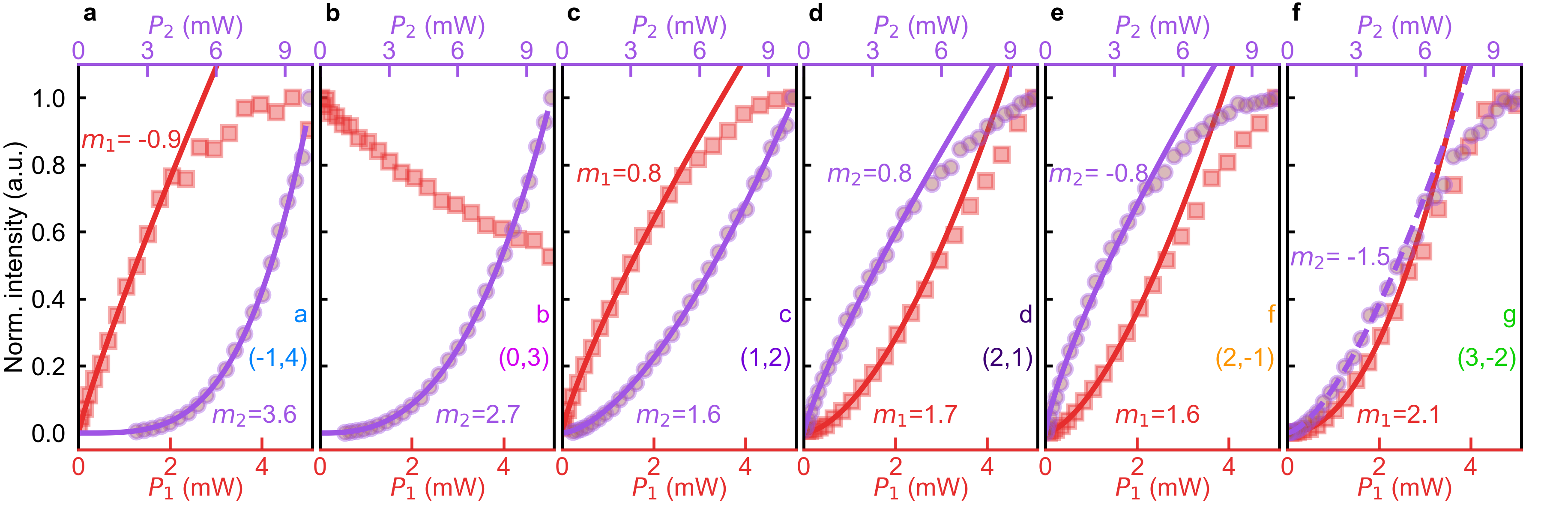} 
	   \caption{ \textbf{Excitation power dependence.} \textbf{a-f}, Intensity of different nonlinear wave mixing signals as a function of 847 nm excitation power $P_1$ and 1280 nm excitation power $P_2$ measured on the bulk MnPSe$_3$ sample. The red and purple markers represent the signal intensities when solely decreasing $P_1$ and $P_2$, respectively. Solid lines are fits to the data. Given the non-perturbative behavior, the red lines in all panels are fitted only with data of $P_1 \le 2$ mW, the purple lines in panel \textbf{d-f} are fitted only with data of $P_2 \le 5$ mW.}
	   \label{fig:2}
    \end{center}
\end{figure}

We studied the intensity of wave mixing peaks as a function of the excitation powers (Fig.~\ref{fig:2}). The power-law fit of data at low excitation powers gives power exponents consistent with expected $|m_1|$ and $|m_2|$ values, albeit slightly lower. At high excitation powers, most of the intensities show saturation feature. The saturation indicates non-perturbative behavior commonly observed in HHG measurements of 2D materials \cite{NaotakaScience2017,LiuNatPhys2017}, but here the saturation is evident even for signals with $|m_1|$ (or $|m_2|$)$\approx 1$, indicating that the giant wave mixing effects strongly perturb the system. Additionally, the THG intensity of 1280 nm laser, which in principle should be independent of the 847 nm laser power, is reduced nearly by half at max power of 847 nm (Fig.~\ref{fig:2}b). This reveals a transient response of the THG to the occurrence of non-degenerate wave mixing near time zero, which we will discuss later.

Among various 2D materials, a typical one with significantly large third-order susceptibility is GaSe (see Table.1 in ref.\cite{AutereAdvMater2018}, Table 4.1.2 in ref.\cite{Boydbook} for summaries of nonlinear susceptibilities of various materials), with $\chi^{(3)}=1.6\times10^{-16} \mathrm{m}^2\mathrm{V}^{-2}$ at 520 nm emission wavelength. In our experiment, different four-wave mixing processes are related to different $\chi^{(3)}$. We compared the THG signals of GaSe and MnPSe$_3$ thin films and estimated the THG $\chi^{(3)}$ of MnPSe$_3$ as shown in Fig.~\ref{fig:1}e (see Supplementary S4 for details). The comparison reveals that the THG efficiency of MnPSe$_3$ film far exceeds that of GaSe. Near 470 nm, MnPSe$_3$ shows $\chi^{(3)}$ values near $1\times10^{-15} \mathrm{m}^2\mathrm{V}^{-2}$, indicating MnPSe$_3$ to be among the group of 2D materials with highest third order susceptibilities. 
We noted that $\chi^{(3)}$ starts to decrease significantly when THG wavelength exceeds 600 nm, which is near the absorption edge in Fig.~\ref{fig:1}a. This hints that the increased density of states above the optical gap resonantly enhances the conversion efficiency for the THG process.

\begin{figure} 
	\begin{center}
		\includegraphics[clip, width=5.5in]{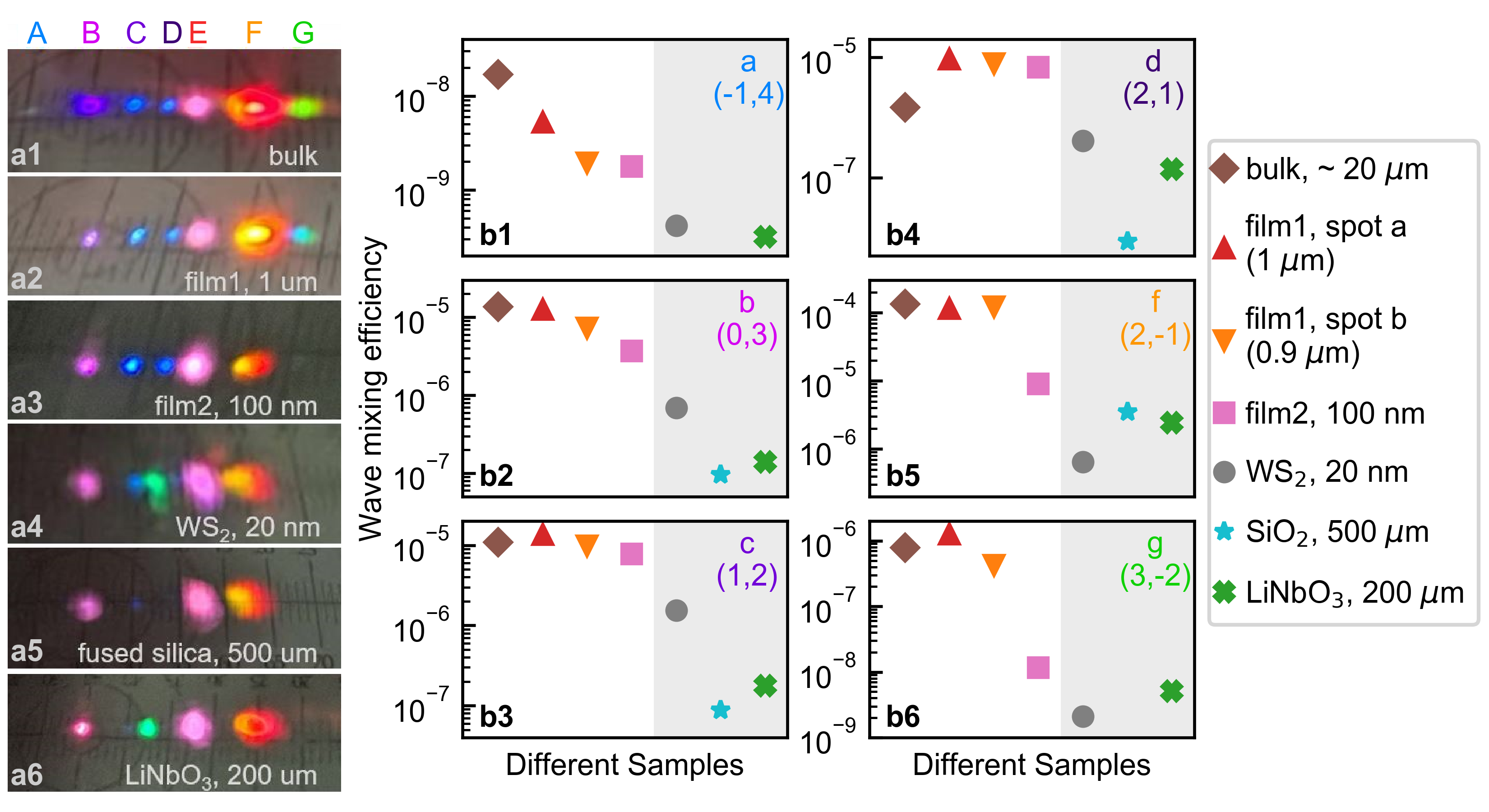} 
		\caption{\textbf{Nonlinear wave mixing efficiency in different samples.} \textbf{a1}-\textbf{a6}, Outgoing beams directly observed on a paper card behind different samples. Note that the green speckles for 3R-WS$_2$ and LiNbO$_3$ samples are the SFG of the 847 nm and 1280 nm lasers. \textbf{b1}-\textbf{b6}, Efficiency of different nonlinear wave mixing processes of different samples shown in \textbf{a1}-\textbf{a6}. The measurements were performed with excitation powers $P_1=5$ mW and $P_2=10$ mW.}
		\label{fig:3}
	\end{center}
\end{figure}

For the bulk sample, beam F (the (2,-1) channel) is strongest among all wave mixing beams, with $\sim2$ $\mu$W power which can be directly measured by a power meter. This yields a four-wave mixing efficiency near the order of $10^{-4}$. As for the six-wave mixing outputs, beam G (the (3,-2) channel) shows the highest efficiency near the order of $10^{-6}$.
We compared the wave mixing efficiency of the bulk sample with other two films of about 1 $\mu$m and 100 nm thicknesses (Fig.~\ref{fig:3}). The difference of wave mixing efficiency is insignificant between the bulk sample and the 1 $\mu$m film.
For the 100 nm film, signal f and g are lowered by about an order of magnitude, while others change slightly. The intensity of nonlinear wave mixing signals is influenced by material thickness mainly in several possible ways, including enhanced absorption of photons with increasing thickness \cite{HendryPRL2010,HongPRX2013} and phase matching effect. 
For the thin films within the order of 100 nm, phase-matching usually plays no role, and the efficiency should grow quadratically with layer numbers in the absence of absorption \cite{HongPRX2013,HongArxiv2023}. For thicker films, the efficiency evolves near-sinusoidally with increasing thickness under phase-mismatching condition. Only when phase-matching is satisfied will the efficiency continue growing quadratically with thickness. The weak dependence of wave mixing efficiency on sample thickness in our measurements indicates the condition of phase-mismatching, which could be optimized to largely improve the wave mixing efficiency in MnPSe$_3$ samples by tuning the phase-matching condition in the future.

We compared the wave mixing efficiency of MnPSe$_3$ with other nonlinear optical materials. LiNbO$_3$ is a typical bulk nonlinear crystal and also commonly used for engineering third-order nonlinear responses through cascaded second-order processes \cite{HunaultOptLett2010,ArahiraOptExp2011,ZhangNpjQI2021}. Fused silica is commonly used for optical elements, and here used as substrates of all thin films. WS$_2$ is a typical 2D vdW material with large nonlinear susceptibility \cite{AutereAdvMater2018}. LiNbO$_3$ bulk and the 3R-WS$_2$ film are both non-centrosymmetric materials with significant second-order nonlinear responses (see the green SFG speckle shown in Fig.~\ref{fig:3}a).
The overall eﬀiciency of different wave mixing processes in LiNbO$_3$ and silica bulks falls below MnPSe$_3$ films by orders of magnitude.
The efficiency in WS$_2$ film (20 nm) is nearly an order of magnitude lower than MnPSe$_3$ film (100 nm). It’s worth noting that, due to the absorption effect, the dependence of wave mixing intensity on the increasing thickness should be below quadratical relation. Besides, the intensity may even show negative dependence on thickness, as we noticed that the THG intensity increased on thinner MnPSe$_3$ films (Supplementary S5). Therefore, the wave mixing eﬀiciency of MnPSe$_3$ is at least comparable to that of WS$_2$ if with the same thickness, rendering it to be among the set of most efficient materials for odd-order nonlinear wave mixing.

\begin{figure*}
    \centering
	   \includegraphics[width=1\textwidth]{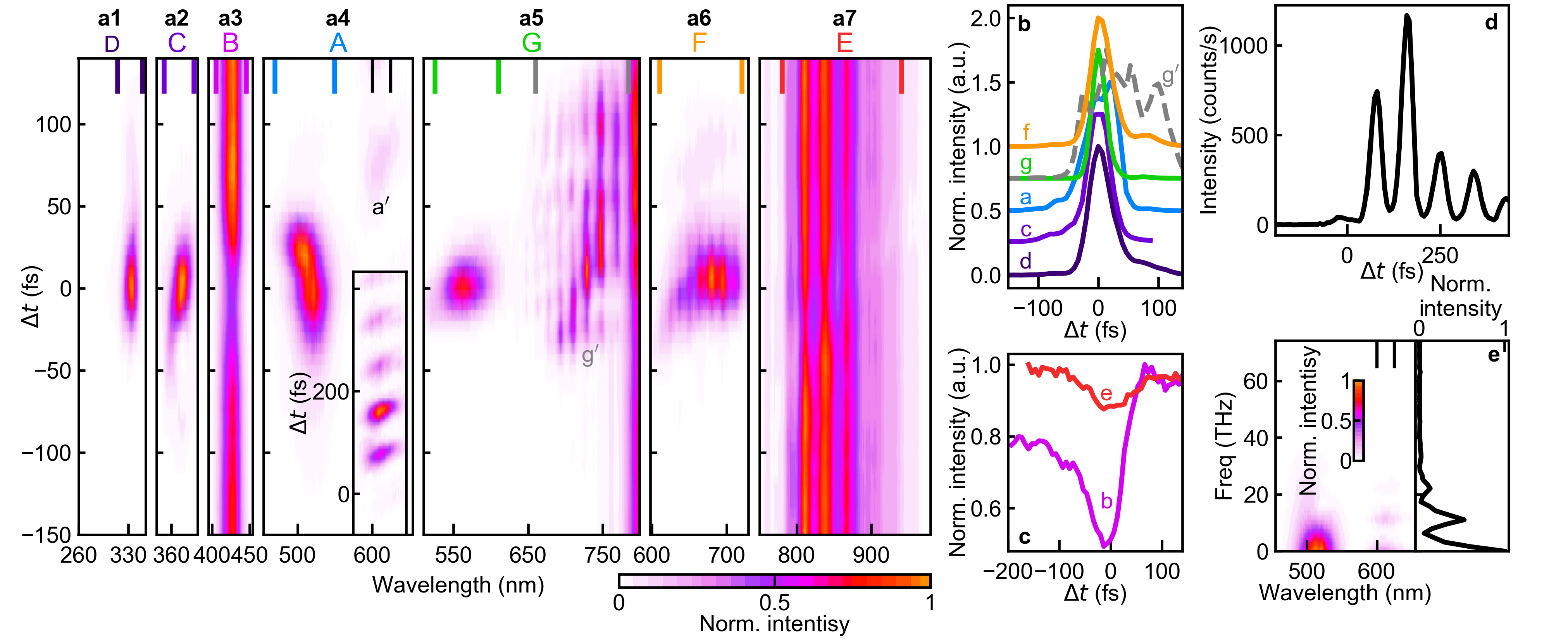} %
	   \caption{\textbf{Temporal dynamics of nonlinear responses.}  \textbf{a1-a7}, Time-dependent spectra of outgoing beam D, C, B, A, G, F, and E, respectively, measured on the bulk MnPSe$_3$ sample. \textbf{b}, Time evolution of integrated intensity of the non-degenerate wave mixing peaks d, c, a, g (and g$^{\prime}$), and f. Each curve is normalized by the max value and vertically shifted for clarity. \textbf{c} Time evolution of integrated intensity of peaks d and e, each normalized by the max value. \textbf{d}, Time evolution of integrated intensity of peak a$^{\prime}$. The integration ranges in panel \textbf{b}, \textbf{c} and \textbf{d} are marked by the colored ticks on top of the axes in \textbf{a1-a7}. \textbf{e}, Left panel, Fast Fourier transformation (FFT) of the time-dependent spectra measured at speckle A presented in panel \textbf{a4}. Right panel, integrated FFT intensities in the range 600-625 nm of the left panels, normalized by the max value.}
	   \label{fig:4}
\end{figure*}

In order to explore the temporal dynamics of the nonlinear responses, we performed time-resolved measurements. Figures \ref{fig:4}a1-a7 show the spectra for beam A-G as a function of $\Delta t=t_{847}-t_{1280}$. The non-degenerate wave mixing signals d, c, a, g and f appear only near $\Delta t=0$. The THG signal b and the 847 nm transmission signal e are present at all time, but peak e is reduced by about 10\% and the THG signal b is reduced nearly by half at $\Delta t =0$ (Fig.~\ref{fig:4}c). The large reduction is likely because that the occurrence of giant non-degenerate nonlinear wave mixing largely consumes the excitation photons. Moreover, the time evolution of signal b and e show dissipative behaviors. Here $\Delta t >0$ ($<0$) means that the 1280 nm pulses arrive before (after) the 847 nm pulses. At negative $\Delta t$, signal b relaxes back to only about 80\% of the original intensity. At positive $\Delta t$, signal e relaxes back to about 95\% of the original intensity. While nonlinear wave mixing is usually regarded as a non-dissipative instantaneous process occurring on the short time scale of the pulse duration, the dissipative behaviors hint that photoinduced carriers are generated along with the multiphoton processes of strong nonlinear wave mixing.

At positive time delays, two unexpected dynamical signals of new energies (denoted as a$^\prime$ and g$^\prime$ in Figs.~\ref{fig:4}a4,a5) appear with the two six-wave mixing signals measured at speckles A and G but with new energies. g$^\prime$ extends longer in time than the wave mixing peak g. a$^\prime$ appears far away from time zero, and shows an oscillatory behavior with 11 THz frequency (Figs.~\ref{fig:4}d,e). It’s also interesting that a$^\prime$ and g$^\prime$ signals vary a lot among different spots on the bulk sample and are absent in the flat films with thicknesses of 1 $\mu$m and 100 nm (Supplementary S6). This indicates that a$^\prime$ and g$^\prime$ signals are not only dependent on the material itself, but extremely sensitive to environmental perturbation, including the thickness and geometric structure of the bulk sample with steps or winkles inside. The exact origin of signal a$^\prime$ and g$^\prime$ deserve further investigation, and here we propose a possible qualitative explanation related to the soliton evolution mechanism. Different from the fundamental soliton whose shape and amplitude remain unchanged with propagation, the high-order soliton exhibits periodical evolution of shape and spectra with time and distance, resulted from the interactions of self-phase modulation and dispersion \cite{DudleyBook,DudleyRevModPhys2006}. It’s worth noting that even slight perturbations to the materials’ dispersive and nonlinear responses would break the symmetry of the evolution, hence the higher-order soliton will undergoes process called soliton fission \cite{DudleyPhysToday2013}, where the initial pulse splits into a train of individual fundamental soliton pulses. As for the newly formed soliton, the effect of inelastic light scattering within the bandwidth of each soliton (also known as intra-pulse Raman scattering) generates the long wavelength side of the continuum, which is also the key to understand the supercontinuum generation process \cite{DudleyRevModPhys2006, DudleyPhysToday2013}.

\section*{Outlook and conclusion}

The MPX$_3$ family contains members with different electronic structure and magnetism \cite{ChittariPRB2016}. We also observed strong wave mixing responses in MnPS$_3$ (Supplementary S7). It is worth investigating whether the nonlinear responses exist in other members and how they depend on the different electronic and magnetic properties. This would also help to theoretically address the exact origin of such large nonlinear responses. Given the layered nature of MPX$_3$ materials, state-of-the-art nano-fabrication and integration techniques, such as vdW engineering (e.g. heterostructuring and moiré structuring) \cite{LiuNature2019,HuangNatNanotech2022}, optical coupling to cavities and waveguides \cite{ZhangNatPhoto2022,XuNatNano2022}, would be promising to enhance the wave mixing eﬀiciency and engineer nonlinear photonic devices.
Our findings not only add a new member to the collection of 2D materials hosting large optical nonlinearities, but also inspire future research in the following perspectives.

Magnetism has always been a research focus of MPX$_3$ materials and greatly modifies the  optical properties \cite{KimPRL2018,ErgecenNatCommu2022,KangNature2020,WangNatMater2021,HwangboNatNanotech2021,BelvinNatCommu2021,NiNatNanotech2021,NiPRL2021,
KimNatCommu2019,ZhangNanoLett2021,ZhangNatPhoto2022,ChuPRL2020,ShanNature2021,WangArxiv2023}. It is worth studying how the nonlinear wave mixing will depend on the magnetic properties. So far, SHG has become powerful for probing magnetic symmetries in 2D materials \cite{NiPRL2021,ChuPRL2020,ShanNature2021,WangArxiv2023,SunNature2019},
but how the high-order nonlinearities will response to magnetic orders remains elusive. The study of high-order nonlinearities in the AFM phases of MPX$_3$ materials will add new insights to such knowledge, with respect to this family as well as magnetic systems in general.
Besides magnetism effects, another intriguing question is whether the wave mixing processes present any excitonic effects when varying excitation laser wavelengths \cite{LowNatMater2017,ReganNatRevMater2022,WangArxiv2023}, given the general fact that the many-body excitons strongly influence the optical properties of MPX$_3$ materials \cite{KangNature2020,HwangboNatNanotech2021,BelvinNatCommu2021}, and resonance to different excitonic states can modify the temporal dynamics of wave mixing processes \cite{BauerNatPhoto2022,ShanNature2021} or activate the precession of magnons \cite{BelvinNatCommu2021}. The multiple $(m_1, m_2)$ wave mixing channels in MPX$_3$ compounds offer unique advantage for clarifying the nonlinear light-matter interactions with different multiphoton excitonic quantum transition paths, while the many-body exciton nature may further enrich the interacting mechanism.

Large odd-order nonlinear susceptibilities in the MPX$_3$ materials can be exploited in other nonlinear optical effects, such as self-phase modulation, saturable absorption and coherent photon conversion, which are all important for nonlinear optical and nano-photonic applications \cite{LangfordNature2011}. Besides, one could try to reach higher-order nonlinear wave mixing effects, such as high-order sideband generation and HHG, which have become an emerging field recently for studying band structure properties \cite{LiuNatPhys2017,CostelloNature2021,YuePRL2022}, field-driven ultrafast electron dynamics \cite{BauerNatPhoto2022,SchubertNatPhoto2014,LangerNature2016,LangerNature2018}, and optical band engineering \cite{UchidaPRL2016}. The endeavor towards higher order wave mixing responses in MPX$_3$ correlated magnetic systems may give new insight to this field.

In all, we discovered giant four-wave and six-wave mixing effects in a wide spectral range from UV to visible in MnPSe$_3$ (and also MnPS$_3$), rendering it to be a model 2D system of large odd-order optical nonlinearities.
Our work highlights the great potential of transition metal phosphorous trichalcogenides family for exploring nonlinear optical processes from both fundamental scientific and applicational point of view. Future research are promising for clarifying the nonlinear light-matter interaction mechanisms in the correlated 2D systems, reaching higher conversion efficiency and engineering new optical devices for nonlinear photonic applications.

\bibliography{Ref_MnPSe3_dressing}

\begin{thebibliography}{50}%
\makeatletter
\providecommand \@ifxundefined [1]{%
 \@ifx{#1\undefined}
}%
\providecommand \@ifnum [1]{%
 \ifnum #1\expandafter \@firstoftwo
 \else \expandafter \@secondoftwo
 \fi
}%
\providecommand \@ifx [1]{%
 \ifx #1\expandafter \@firstoftwo
 \else \expandafter \@secondoftwo
 \fi
}%
\providecommand \natexlab [1]{#1}%
\providecommand \enquote  [1]{``#1''}%
\providecommand \bibnamefont  [1]{#1}%
\providecommand \bibfnamefont [1]{#1}%
\providecommand \citenamefont [1]{#1}%
\providecommand \href@noop [0]{\@secondoftwo}%
\providecommand \href [0]{\begingroup \@sanitize@url \@href}%
\providecommand \@href[1]{\@@startlink{#1}\@@href}%
\providecommand \@@href[1]{\endgroup#1\@@endlink}%
\providecommand \@sanitize@url [0]{\catcode `\\12\catcode `\$12\catcode
  `\&12\catcode `\#12\catcode `\^12\catcode `\_12\catcode `\%12\relax}%
\providecommand \@@startlink[1]{}%
\providecommand \@@endlink[0]{}%
\providecommand \url  [0]{\begingroup\@sanitize@url \@url }%
\providecommand \@url [1]{\endgroup\@href {#1}{\urlprefix }}%
\providecommand \urlprefix  [0]{URL }%
\providecommand \Eprint [0]{\href }%
\providecommand \doibase [0]{https://doi.org/}%
\providecommand \selectlanguage [0]{\@gobble}%
\providecommand \bibinfo  [0]{\@secondoftwo}%
\providecommand \bibfield  [0]{\@secondoftwo}%
\providecommand \translation [1]{[#1]}%
\providecommand \BibitemOpen [0]{}%
\providecommand \bibitemStop [0]{}%
\providecommand \bibitemNoStop [0]{.\EOS\space}%
\providecommand \EOS [0]{\spacefactor3000\relax}%
\providecommand \BibitemShut  [1]{\csname bibitem#1\endcsname}%
\let\auto@bib@innerbib\@empty
\bibitem [{\citenamefont {Boyd}(2007)}]{Boydbook}%
  \BibitemOpen
  \bibfield  {author} {\bibinfo {author} {\bibfnamefont {R.~W.}\ \bibnamefont
  {Boyd}},\ }\href@noop {} {\emph {\bibinfo {title} {Nonlinear optics}}}\
  (\bibinfo  {publisher} {Academic Press},\ \bibinfo {year} {2007})\BibitemShut
  {NoStop}%
\bibitem [{\citenamefont {Shen}(1984)}]{Shenbook}%
  \BibitemOpen
  \bibfield  {author} {\bibinfo {author} {\bibfnamefont {Y.}~\bibnamefont
  {Shen}},\ }\href@noop {} {\emph {\bibinfo {title} {The principles of
  nonlinear optics}}}\ (\bibinfo  {publisher} {Wiley Press},\ \bibinfo {year}
  {1984})\BibitemShut {NoStop}%
\bibitem [{\citenamefont {Duelli}\ \emph {et~al.}(2000)\citenamefont {Duelli},
  \citenamefont {Montemezzani}, \citenamefont {Zgonik},\ and\ \citenamefont
  {Günter}}]{Duellibook}%
  \BibitemOpen
  \bibfield  {author} {\bibinfo {author} {\bibfnamefont {M.}~\bibnamefont
  {Duelli}}, \bibinfo {author} {\bibfnamefont {G.}~\bibnamefont
  {Montemezzani}}, \bibinfo {author} {\bibfnamefont {M.}~\bibnamefont
  {Zgonik}},\ and\ \bibinfo {author} {\bibfnamefont {P.}~\bibnamefont
  {Günter}},\ }\href@noop {} {\emph {\bibinfo {title} {Nonlinear Optical
  Effects and Materials}}}\ (\bibinfo  {publisher} {Springer},\ \bibinfo {year}
  {2000})\BibitemShut {NoStop}%
\bibitem [{\citenamefont {Keller}(2003)}]{KellerNature2003}%
  \BibitemOpen
  \bibfield  {author} {\bibinfo {author} {\bibfnamefont {U.}~\bibnamefont
  {Keller}},\ }\href {https://doi.org/10.1038/nature01938} {\bibfield
  {journal} {\bibinfo  {journal} {Nature}\ }\textbf {\bibinfo {volume} {424}},\
  \bibinfo {pages} {831} (\bibinfo {year} {2003})}\BibitemShut {NoStop}%
\bibitem [{\citenamefont {Garmire}(2013)}]{GarmireOptExp2013}%
  \BibitemOpen
  \bibfield  {author} {\bibinfo {author} {\bibfnamefont {E.}~\bibnamefont
  {Garmire}},\ }\href {https://doi.org/10.1364/OE.21.030532} {\bibfield
  {journal} {\bibinfo  {journal} {Opt. Express}\ }\textbf {\bibinfo {volume}
  {21}},\ \bibinfo {pages} {30532} (\bibinfo {year} {2013})}\BibitemShut
  {NoStop}%
\bibitem [{\citenamefont {Autere}\ \emph {et~al.}(2018)\citenamefont {Autere},
  \citenamefont {Jussila}, \citenamefont {Dai}, \citenamefont {Wang},
  \citenamefont {Lipsanen},\ and\ \citenamefont {Sun}}]{AutereAdvMater2018}%
  \BibitemOpen
  \bibfield  {author} {\bibinfo {author} {\bibfnamefont {A.}~\bibnamefont
  {Autere}}, \bibinfo {author} {\bibfnamefont {H.}~\bibnamefont {Jussila}},
  \bibinfo {author} {\bibfnamefont {Y.}~\bibnamefont {Dai}}, \bibinfo {author}
  {\bibfnamefont {Y.}~\bibnamefont {Wang}}, \bibinfo {author} {\bibfnamefont
  {H.}~\bibnamefont {Lipsanen}},\ and\ \bibinfo {author} {\bibfnamefont
  {Z.}~\bibnamefont {Sun}},\ }\href
  {https://doi.org/https://doi.org/10.1002/adma.201705963} {\bibfield
  {journal} {\bibinfo  {journal} {Advanced Materials}\ }\textbf {\bibinfo
  {volume} {30}},\ \bibinfo {pages} {1705963} (\bibinfo {year}
  {2018})}\BibitemShut {NoStop}%
\bibitem [{\citenamefont {Chang}\ \emph {et~al.}(2014)\citenamefont {Chang},
  \citenamefont {Vuleti{\'{c}}},\ and\ \citenamefont
  {Lukin}}]{ChangNatPhoto2014}%
  \BibitemOpen
  \bibfield  {author} {\bibinfo {author} {\bibfnamefont {D.~E.}\ \bibnamefont
  {Chang}}, \bibinfo {author} {\bibfnamefont {V.}~\bibnamefont
  {Vuleti{\'{c}}}},\ and\ \bibinfo {author} {\bibfnamefont {M.~D.}\
  \bibnamefont {Lukin}},\ }\href {https://doi.org/10.1038/nphoton.2014.192}
  {\bibfield  {journal} {\bibinfo  {journal} {Nature Photonics}\ }\textbf
  {\bibinfo {volume} {8}},\ \bibinfo {pages} {685} (\bibinfo {year}
  {2014})}\BibitemShut {NoStop}%
\bibitem [{\citenamefont {Elshaari}\ \emph {et~al.}(2020)\citenamefont
  {Elshaari}, \citenamefont {Pernice}, \citenamefont {Srinivasan},
  \citenamefont {Benson},\ and\ \citenamefont
  {Zwiller}}]{ElshaariNatPhoto2020}%
  \BibitemOpen
  \bibfield  {author} {\bibinfo {author} {\bibfnamefont {A.~W.}\ \bibnamefont
  {Elshaari}}, \bibinfo {author} {\bibfnamefont {W.}~\bibnamefont {Pernice}},
  \bibinfo {author} {\bibfnamefont {K.}~\bibnamefont {Srinivasan}}, \bibinfo
  {author} {\bibfnamefont {O.}~\bibnamefont {Benson}},\ and\ \bibinfo {author}
  {\bibfnamefont {V.}~\bibnamefont {Zwiller}},\ }\href
  {https://doi.org/10.1038/s41566-020-0609-x} {\bibfield  {journal} {\bibinfo
  {journal} {Nature Photonics}\ }\textbf {\bibinfo {volume} {14}},\ \bibinfo
  {pages} {285} (\bibinfo {year} {2020})}\BibitemShut {NoStop}%
\bibitem [{\citenamefont {Liu}\ \emph {et~al.}(2019)\citenamefont {Liu},
  \citenamefont {Huang},\ and\ \citenamefont {Duan}}]{LiuNature2019}%
  \BibitemOpen
  \bibfield  {author} {\bibinfo {author} {\bibfnamefont {Y.}~\bibnamefont
  {Liu}}, \bibinfo {author} {\bibfnamefont {Y.}~\bibnamefont {Huang}},\ and\
  \bibinfo {author} {\bibfnamefont {X.}~\bibnamefont {Duan}},\ }\href
  {https://doi.org/10.1038/s41586-019-1013-x} {\bibfield  {journal} {\bibinfo
  {journal} {Nature}\ }\textbf {\bibinfo {volume} {567}},\ \bibinfo {pages}
  {323} (\bibinfo {year} {2019})}\BibitemShut {NoStop}%
\bibitem [{\citenamefont {Guo}\ \emph {et~al.}(2019)\citenamefont {Guo},
  \citenamefont {Xiao}, \citenamefont {Wang},\ and\ \citenamefont
  {Zhang}}]{GuoLaserPhoRev2019}%
  \BibitemOpen
  \bibfield  {author} {\bibinfo {author} {\bibfnamefont {B.}~\bibnamefont
  {Guo}}, \bibinfo {author} {\bibfnamefont {Q.-l.}\ \bibnamefont {Xiao}},
  \bibinfo {author} {\bibfnamefont {S.-h.}\ \bibnamefont {Wang}},\ and\
  \bibinfo {author} {\bibfnamefont {H.}~\bibnamefont {Zhang}},\ }\href
  {https://doi.org/https://doi.org/10.1002/lpor.201800327} {\bibfield
  {journal} {\bibinfo  {journal} {Laser \& Photonics Reviews}\ }\textbf
  {\bibinfo {volume} {13}},\ \bibinfo {pages} {1800327} (\bibinfo {year}
  {2019})}\BibitemShut {NoStop}%
\bibitem [{\citenamefont {Hunault}\ \emph {et~al.}(2010)\citenamefont
  {Hunault}, \citenamefont {Takesue}, \citenamefont {Tadanaga}, \citenamefont
  {Nishida},\ and\ \citenamefont {Asobe}}]{HunaultOptLett2010}%
  \BibitemOpen
  \bibfield  {author} {\bibinfo {author} {\bibfnamefont {M.}~\bibnamefont
  {Hunault}}, \bibinfo {author} {\bibfnamefont {H.}~\bibnamefont {Takesue}},
  \bibinfo {author} {\bibfnamefont {O.}~\bibnamefont {Tadanaga}}, \bibinfo
  {author} {\bibfnamefont {Y.}~\bibnamefont {Nishida}},\ and\ \bibinfo {author}
  {\bibfnamefont {M.}~\bibnamefont {Asobe}},\ }\href
  {https://doi.org/10.1364/OL.35.001239} {\bibfield  {journal} {\bibinfo
  {journal} {Opt. Lett.}\ }\textbf {\bibinfo {volume} {35}},\ \bibinfo {pages}
  {1239} (\bibinfo {year} {2010})}\BibitemShut {NoStop}%
\bibitem [{\citenamefont {Arahira}\ \emph {et~al.}(2011)\citenamefont
  {Arahira}, \citenamefont {Namekata}, \citenamefont {Kishimoto}, \citenamefont
  {Yaegashi},\ and\ \citenamefont {Inoue}}]{ArahiraOptExp2011}%
  \BibitemOpen
  \bibfield  {author} {\bibinfo {author} {\bibfnamefont {S.}~\bibnamefont
  {Arahira}}, \bibinfo {author} {\bibfnamefont {N.}~\bibnamefont {Namekata}},
  \bibinfo {author} {\bibfnamefont {T.}~\bibnamefont {Kishimoto}}, \bibinfo
  {author} {\bibfnamefont {H.}~\bibnamefont {Yaegashi}},\ and\ \bibinfo
  {author} {\bibfnamefont {S.}~\bibnamefont {Inoue}},\ }\href
  {https://doi.org/10.1364/OE.19.016032} {\bibfield  {journal} {\bibinfo
  {journal} {Opt. Express}\ }\textbf {\bibinfo {volume} {19}},\ \bibinfo
  {pages} {16032} (\bibinfo {year} {2011})}\BibitemShut {NoStop}%
\bibitem [{\citenamefont {Zhang}\ \emph
  {et~al.}(2021{\natexlab{a}})\citenamefont {Zhang}, \citenamefont {Yuan},
  \citenamefont {Shen}, \citenamefont {Yu}, \citenamefont {Zhang},
  \citenamefont {Wang}, \citenamefont {Li}, \citenamefont {Wang}, \citenamefont
  {Deng}, \citenamefont {Wang}, \citenamefont {You}, \citenamefont {Wang},
  \citenamefont {Song}, \citenamefont {Guo},\ and\ \citenamefont
  {Zhou}}]{ZhangNpjQI2021}%
  \BibitemOpen
  \bibfield  {author} {\bibinfo {author} {\bibfnamefont {Z.}~\bibnamefont
  {Zhang}}, \bibinfo {author} {\bibfnamefont {C.}~\bibnamefont {Yuan}},
  \bibinfo {author} {\bibfnamefont {S.}~\bibnamefont {Shen}}, \bibinfo {author}
  {\bibfnamefont {H.}~\bibnamefont {Yu}}, \bibinfo {author} {\bibfnamefont
  {R.}~\bibnamefont {Zhang}}, \bibinfo {author} {\bibfnamefont
  {H.}~\bibnamefont {Wang}}, \bibinfo {author} {\bibfnamefont {H.}~\bibnamefont
  {Li}}, \bibinfo {author} {\bibfnamefont {Y.}~\bibnamefont {Wang}}, \bibinfo
  {author} {\bibfnamefont {G.}~\bibnamefont {Deng}}, \bibinfo {author}
  {\bibfnamefont {Z.}~\bibnamefont {Wang}}, \bibinfo {author} {\bibfnamefont
  {L.}~\bibnamefont {You}}, \bibinfo {author} {\bibfnamefont {Z.}~\bibnamefont
  {Wang}}, \bibinfo {author} {\bibfnamefont {H.}~\bibnamefont {Song}}, \bibinfo
  {author} {\bibfnamefont {G.}~\bibnamefont {Guo}},\ and\ \bibinfo {author}
  {\bibfnamefont {Q.}~\bibnamefont {Zhou}},\ }\href
  {https://doi.org/10.1038/s41534-021-00462-7} {\bibfield  {journal} {\bibinfo
  {journal} {npj Quantum Information}\ }\textbf {\bibinfo {volume} {7}},\
  \bibinfo {pages} {123} (\bibinfo {year} {2021}{\natexlab{a}})}\BibitemShut
  {NoStop}%
\bibitem [{\citenamefont {Hendry}\ \emph {et~al.}(2010)\citenamefont {Hendry},
  \citenamefont {Hale}, \citenamefont {Moger}, \citenamefont {Savchenko},\ and\
  \citenamefont {Mikhailov}}]{HendryPRL2010}%
  \BibitemOpen
  \bibfield  {author} {\bibinfo {author} {\bibfnamefont {E.}~\bibnamefont
  {Hendry}}, \bibinfo {author} {\bibfnamefont {P.~J.}\ \bibnamefont {Hale}},
  \bibinfo {author} {\bibfnamefont {J.}~\bibnamefont {Moger}}, \bibinfo
  {author} {\bibfnamefont {A.~K.}\ \bibnamefont {Savchenko}},\ and\ \bibinfo
  {author} {\bibfnamefont {S.~A.}\ \bibnamefont {Mikhailov}},\ }\href
  {https://doi.org/10.1103/PhysRevLett.105.097401} {\bibfield  {journal}
  {\bibinfo  {journal} {Phys. Rev. Lett.}\ }\textbf {\bibinfo {volume} {105}},\
  \bibinfo {pages} {097401} (\bibinfo {year} {2010})}\BibitemShut {NoStop}%
\bibitem [{\citenamefont {Hong}\ \emph {et~al.}(2013)\citenamefont {Hong},
  \citenamefont {Dadap}, \citenamefont {Petrone}, \citenamefont {Yeh},
  \citenamefont {Hone},\ and\ \citenamefont {Osgood}}]{HongPRX2013}%
  \BibitemOpen
  \bibfield  {author} {\bibinfo {author} {\bibfnamefont {S.-Y.}\ \bibnamefont
  {Hong}}, \bibinfo {author} {\bibfnamefont {J.~I.}\ \bibnamefont {Dadap}},
  \bibinfo {author} {\bibfnamefont {N.}~\bibnamefont {Petrone}}, \bibinfo
  {author} {\bibfnamefont {P.-C.}\ \bibnamefont {Yeh}}, \bibinfo {author}
  {\bibfnamefont {J.}~\bibnamefont {Hone}},\ and\ \bibinfo {author}
  {\bibfnamefont {R.~M.}\ \bibnamefont {Osgood}},\ }\href
  {https://doi.org/10.1103/PhysRevX.3.021014} {\bibfield  {journal} {\bibinfo
  {journal} {Phys. Rev. X}\ }\textbf {\bibinfo {volume} {3}},\ \bibinfo {pages}
  {021014} (\bibinfo {year} {2013})}\BibitemShut {NoStop}%
\bibitem [{\citenamefont {Hernandez-Rueda}\ \emph {et~al.}(2021)\citenamefont
  {Hernandez-Rueda}, \citenamefont {Noordam}, \citenamefont {Komen},\ and\
  \citenamefont {Kuipers}}]{RuedaACSPhoto2021}%
  \BibitemOpen
  \bibfield  {author} {\bibinfo {author} {\bibfnamefont {J.}~\bibnamefont
  {Hernandez-Rueda}}, \bibinfo {author} {\bibfnamefont {M.~L.}\ \bibnamefont
  {Noordam}}, \bibinfo {author} {\bibfnamefont {I.}~\bibnamefont {Komen}},\
  and\ \bibinfo {author} {\bibfnamefont {L.}~\bibnamefont {Kuipers}},\ }\href
  {https://doi.org/10.1021/acsphotonics.0c01567} {\bibfield  {journal}
  {\bibinfo  {journal} {ACS Photonics}\ }\textbf {\bibinfo {volume} {8}},\
  \bibinfo {pages} {550} (\bibinfo {year} {2021})}\BibitemShut {NoStop}%
\bibitem [{\citenamefont {Bauer}\ \emph {et~al.}(2022)\citenamefont {Bauer},
  \citenamefont {Chen}, \citenamefont {Wilhelm}, \citenamefont {Watanabe},
  \citenamefont {Taniguchi}, \citenamefont {Bange}, \citenamefont {Lupton},\
  and\ \citenamefont {Lin}}]{BauerNatPhoto2022}%
  \BibitemOpen
  \bibfield  {author} {\bibinfo {author} {\bibfnamefont {J.~M.}\ \bibnamefont
  {Bauer}}, \bibinfo {author} {\bibfnamefont {L.}~\bibnamefont {Chen}},
  \bibinfo {author} {\bibfnamefont {P.}~\bibnamefont {Wilhelm}}, \bibinfo
  {author} {\bibfnamefont {K.}~\bibnamefont {Watanabe}}, \bibinfo {author}
  {\bibfnamefont {T.}~\bibnamefont {Taniguchi}}, \bibinfo {author}
  {\bibfnamefont {S.}~\bibnamefont {Bange}}, \bibinfo {author} {\bibfnamefont
  {J.~M.}\ \bibnamefont {Lupton}},\ and\ \bibinfo {author} {\bibfnamefont
  {K.-Q.}\ \bibnamefont {Lin}},\ }\href
  {https://doi.org/10.1038/s41566-022-01080-1} {\bibfield  {journal} {\bibinfo
  {journal} {Nature Photonics}\ }\textbf {\bibinfo {volume} {16}},\ \bibinfo
  {pages} {777} (\bibinfo {year} {2022})}\BibitemShut {NoStop}%
\bibitem [{\citenamefont {Kim}\ \emph {et~al.}(2018)\citenamefont {Kim},
  \citenamefont {Kim}, \citenamefont {Sandilands}, \citenamefont {Sinn},
  \citenamefont {Lee}, \citenamefont {Son}, \citenamefont {Lee}, \citenamefont
  {Choi}, \citenamefont {Kim}, \citenamefont {Park}, \citenamefont {Jeon},
  \citenamefont {Kim}, \citenamefont {Park}, \citenamefont {Park},
  \citenamefont {Moon},\ and\ \citenamefont {Noh}}]{KimPRL2018}%
  \BibitemOpen
  \bibfield  {author} {\bibinfo {author} {\bibfnamefont {S.~Y.}\ \bibnamefont
  {Kim}}, \bibinfo {author} {\bibfnamefont {T.~Y.}\ \bibnamefont {Kim}},
  \bibinfo {author} {\bibfnamefont {L.~J.}\ \bibnamefont {Sandilands}},
  \bibinfo {author} {\bibfnamefont {S.}~\bibnamefont {Sinn}}, \bibinfo {author}
  {\bibfnamefont {M.-C.}\ \bibnamefont {Lee}}, \bibinfo {author} {\bibfnamefont
  {J.}~\bibnamefont {Son}}, \bibinfo {author} {\bibfnamefont {S.}~\bibnamefont
  {Lee}}, \bibinfo {author} {\bibfnamefont {K.-Y.}\ \bibnamefont {Choi}},
  \bibinfo {author} {\bibfnamefont {W.}~\bibnamefont {Kim}}, \bibinfo {author}
  {\bibfnamefont {B.-G.}\ \bibnamefont {Park}}, \bibinfo {author}
  {\bibfnamefont {C.}~\bibnamefont {Jeon}}, \bibinfo {author} {\bibfnamefont
  {H.-D.}\ \bibnamefont {Kim}}, \bibinfo {author} {\bibfnamefont {C.-H.}\
  \bibnamefont {Park}}, \bibinfo {author} {\bibfnamefont {J.-G.}\ \bibnamefont
  {Park}}, \bibinfo {author} {\bibfnamefont {S.~J.}\ \bibnamefont {Moon}},\
  and\ \bibinfo {author} {\bibfnamefont {T.~W.}\ \bibnamefont {Noh}},\ }\href
  {https://doi.org/10.1103/PhysRevLett.120.136402} {\bibfield  {journal}
  {\bibinfo  {journal} {Phys. Rev. Lett.}\ }\textbf {\bibinfo {volume} {120}},\
  \bibinfo {pages} {136402} (\bibinfo {year} {2018})}\BibitemShut {NoStop}%
\bibitem [{\citenamefont {Erge{\c{c}}en}\ \emph {et~al.}(2022)\citenamefont
  {Erge{\c{c}}en}, \citenamefont {Ilyas}, \citenamefont {Mao}, \citenamefont
  {Po}, \citenamefont {Yilmaz}, \citenamefont {Kim}, \citenamefont {Park},
  \citenamefont {Senthil},\ and\ \citenamefont {Gedik}}]{ErgecenNatCommu2022}%
  \BibitemOpen
  \bibfield  {author} {\bibinfo {author} {\bibfnamefont {E.}~\bibnamefont
  {Erge{\c{c}}en}}, \bibinfo {author} {\bibfnamefont {B.}~\bibnamefont
  {Ilyas}}, \bibinfo {author} {\bibfnamefont {D.}~\bibnamefont {Mao}}, \bibinfo
  {author} {\bibfnamefont {H.~C.}\ \bibnamefont {Po}}, \bibinfo {author}
  {\bibfnamefont {M.~B.}\ \bibnamefont {Yilmaz}}, \bibinfo {author}
  {\bibfnamefont {J.}~\bibnamefont {Kim}}, \bibinfo {author} {\bibfnamefont
  {J.-G.}\ \bibnamefont {Park}}, \bibinfo {author} {\bibfnamefont
  {T.}~\bibnamefont {Senthil}},\ and\ \bibinfo {author} {\bibfnamefont
  {N.}~\bibnamefont {Gedik}},\ }\href
  {https://doi.org/10.1038/s41467-021-27741-3} {\bibfield  {journal} {\bibinfo
  {journal} {Nature Communications}\ }\textbf {\bibinfo {volume} {13}},\
  \bibinfo {pages} {98} (\bibinfo {year} {2022})}\BibitemShut {NoStop}%
\bibitem [{\citenamefont {Wang}\ \emph {et~al.}(2021)\citenamefont {Wang},
  \citenamefont {Cao}, \citenamefont {Lu}, \citenamefont {Cohen}, \citenamefont
  {Kitadai}, \citenamefont {Li}, \citenamefont {Tan}, \citenamefont {Wilson},
  \citenamefont {Lui}, \citenamefont {Smirnov}, \citenamefont {Sharifzadeh},\
  and\ \citenamefont {Ling}}]{WangNatMater2021}%
  \BibitemOpen
  \bibfield  {author} {\bibinfo {author} {\bibfnamefont {X.}~\bibnamefont
  {Wang}}, \bibinfo {author} {\bibfnamefont {J.}~\bibnamefont {Cao}}, \bibinfo
  {author} {\bibfnamefont {Z.}~\bibnamefont {Lu}}, \bibinfo {author}
  {\bibfnamefont {A.}~\bibnamefont {Cohen}}, \bibinfo {author} {\bibfnamefont
  {H.}~\bibnamefont {Kitadai}}, \bibinfo {author} {\bibfnamefont
  {T.}~\bibnamefont {Li}}, \bibinfo {author} {\bibfnamefont {Q.}~\bibnamefont
  {Tan}}, \bibinfo {author} {\bibfnamefont {M.}~\bibnamefont {Wilson}},
  \bibinfo {author} {\bibfnamefont {C.~H.}\ \bibnamefont {Lui}}, \bibinfo
  {author} {\bibfnamefont {D.}~\bibnamefont {Smirnov}}, \bibinfo {author}
  {\bibfnamefont {S.}~\bibnamefont {Sharifzadeh}},\ and\ \bibinfo {author}
  {\bibfnamefont {X.}~\bibnamefont {Ling}},\ }\href
  {https://doi.org/10.1038/s41563-021-00968-7} {\bibfield  {journal} {\bibinfo
  {journal} {Nature Materials}\ }\textbf {\bibinfo {volume} {20}},\ \bibinfo
  {pages} {964} (\bibinfo {year} {2021})}\BibitemShut {NoStop}%
\bibitem [{\citenamefont {Kang}\ \emph {et~al.}(2020)\citenamefont {Kang},
  \citenamefont {Kim}, \citenamefont {Kim}, \citenamefont {Kim}, \citenamefont
  {Sim}, \citenamefont {Lee}, \citenamefont {Lee}, \citenamefont {Park},
  \citenamefont {Yun}, \citenamefont {Kim}, \citenamefont {Nag}, \citenamefont
  {Walters}, \citenamefont {Garcia-Fernandez}, \citenamefont {Li},
  \citenamefont {Chapon}, \citenamefont {Zhou}, \citenamefont {Son},
  \citenamefont {Kim}, \citenamefont {Cheong},\ and\ \citenamefont
  {Park}}]{KangNature2020}%
  \BibitemOpen
  \bibfield  {author} {\bibinfo {author} {\bibfnamefont {S.}~\bibnamefont
  {Kang}}, \bibinfo {author} {\bibfnamefont {K.}~\bibnamefont {Kim}}, \bibinfo
  {author} {\bibfnamefont {B.~H.}\ \bibnamefont {Kim}}, \bibinfo {author}
  {\bibfnamefont {J.}~\bibnamefont {Kim}}, \bibinfo {author} {\bibfnamefont
  {K.~I.}\ \bibnamefont {Sim}}, \bibinfo {author} {\bibfnamefont {J.-U.}\
  \bibnamefont {Lee}}, \bibinfo {author} {\bibfnamefont {S.}~\bibnamefont
  {Lee}}, \bibinfo {author} {\bibfnamefont {K.}~\bibnamefont {Park}}, \bibinfo
  {author} {\bibfnamefont {S.}~\bibnamefont {Yun}}, \bibinfo {author}
  {\bibfnamefont {T.}~\bibnamefont {Kim}}, \bibinfo {author} {\bibfnamefont
  {A.}~\bibnamefont {Nag}}, \bibinfo {author} {\bibfnamefont {A.}~\bibnamefont
  {Walters}}, \bibinfo {author} {\bibfnamefont {M.}~\bibnamefont
  {Garcia-Fernandez}}, \bibinfo {author} {\bibfnamefont {J.}~\bibnamefont
  {Li}}, \bibinfo {author} {\bibfnamefont {L.}~\bibnamefont {Chapon}}, \bibinfo
  {author} {\bibfnamefont {K.-J.}\ \bibnamefont {Zhou}}, \bibinfo {author}
  {\bibfnamefont {Y.-W.}\ \bibnamefont {Son}}, \bibinfo {author} {\bibfnamefont
  {J.~H.}\ \bibnamefont {Kim}}, \bibinfo {author} {\bibfnamefont
  {H.}~\bibnamefont {Cheong}},\ and\ \bibinfo {author} {\bibfnamefont {J.-G.}\
  \bibnamefont {Park}},\ }\href {https://doi.org/10.1038/s41586-020-2520-5}
  {\bibfield  {journal} {\bibinfo  {journal} {Nature}\ }\textbf {\bibinfo
  {volume} {583}},\ \bibinfo {pages} {785} (\bibinfo {year}
  {2020})}\BibitemShut {NoStop}%
\bibitem [{\citenamefont {Hwangbo}\ \emph {et~al.}(2021)\citenamefont
  {Hwangbo}, \citenamefont {Zhang}, \citenamefont {Jiang}, \citenamefont
  {Wang}, \citenamefont {Fonseca}, \citenamefont {Wang}, \citenamefont
  {Diederich}, \citenamefont {Gamelin}, \citenamefont {Xiao}, \citenamefont
  {Chu}, \citenamefont {Yao},\ and\ \citenamefont
  {Xu}}]{HwangboNatNanotech2021}%
  \BibitemOpen
  \bibfield  {author} {\bibinfo {author} {\bibfnamefont {K.}~\bibnamefont
  {Hwangbo}}, \bibinfo {author} {\bibfnamefont {Q.}~\bibnamefont {Zhang}},
  \bibinfo {author} {\bibfnamefont {Q.}~\bibnamefont {Jiang}}, \bibinfo
  {author} {\bibfnamefont {Y.}~\bibnamefont {Wang}}, \bibinfo {author}
  {\bibfnamefont {J.}~\bibnamefont {Fonseca}}, \bibinfo {author} {\bibfnamefont
  {C.}~\bibnamefont {Wang}}, \bibinfo {author} {\bibfnamefont {G.~M.}\
  \bibnamefont {Diederich}}, \bibinfo {author} {\bibfnamefont {D.~R.}\
  \bibnamefont {Gamelin}}, \bibinfo {author} {\bibfnamefont {D.}~\bibnamefont
  {Xiao}}, \bibinfo {author} {\bibfnamefont {J.-H.}\ \bibnamefont {Chu}},
  \bibinfo {author} {\bibfnamefont {W.}~\bibnamefont {Yao}},\ and\ \bibinfo
  {author} {\bibfnamefont {X.}~\bibnamefont {Xu}},\ }\href
  {https://doi.org/10.1038/s41565-021-00873-9} {\bibfield  {journal} {\bibinfo
  {journal} {Nature Nanotechnology}\ }\textbf {\bibinfo {volume} {16}},\
  \bibinfo {pages} {655} (\bibinfo {year} {2021})}\BibitemShut {NoStop}%
\bibitem [{\citenamefont {Belvin}\ \emph {et~al.}(2021)\citenamefont {Belvin},
  \citenamefont {Baldini}, \citenamefont {Ozel}, \citenamefont {Mao},
  \citenamefont {Po}, \citenamefont {Allington}, \citenamefont {Son},
  \citenamefont {Kim}, \citenamefont {Kim}, \citenamefont {Hwang},
  \citenamefont {Kim}, \citenamefont {Park}, \citenamefont {Senthil},\ and\
  \citenamefont {Gedik}}]{BelvinNatCommu2021}%
  \BibitemOpen
  \bibfield  {author} {\bibinfo {author} {\bibfnamefont {C.~A.}\ \bibnamefont
  {Belvin}}, \bibinfo {author} {\bibfnamefont {E.}~\bibnamefont {Baldini}},
  \bibinfo {author} {\bibfnamefont {I.~O.}\ \bibnamefont {Ozel}}, \bibinfo
  {author} {\bibfnamefont {D.}~\bibnamefont {Mao}}, \bibinfo {author}
  {\bibfnamefont {H.~C.}\ \bibnamefont {Po}}, \bibinfo {author} {\bibfnamefont
  {C.~J.}\ \bibnamefont {Allington}}, \bibinfo {author} {\bibfnamefont
  {S.}~\bibnamefont {Son}}, \bibinfo {author} {\bibfnamefont {B.~H.}\
  \bibnamefont {Kim}}, \bibinfo {author} {\bibfnamefont {J.}~\bibnamefont
  {Kim}}, \bibinfo {author} {\bibfnamefont {I.}~\bibnamefont {Hwang}}, \bibinfo
  {author} {\bibfnamefont {J.~H.}\ \bibnamefont {Kim}}, \bibinfo {author}
  {\bibfnamefont {J.-G.}\ \bibnamefont {Park}}, \bibinfo {author}
  {\bibfnamefont {T.}~\bibnamefont {Senthil}},\ and\ \bibinfo {author}
  {\bibfnamefont {N.}~\bibnamefont {Gedik}},\ }\href
  {https://doi.org/10.1038/s41467-021-25164-8} {\bibfield  {journal} {\bibinfo
  {journal} {Nature Communications}\ }\textbf {\bibinfo {volume} {12}},\
  \bibinfo {pages} {4837} (\bibinfo {year} {2021})}\BibitemShut {NoStop}%
\bibitem [{\citenamefont {Ni}\ \emph {et~al.}(2021)\citenamefont {Ni},
  \citenamefont {Haglund}, \citenamefont {Wang}, \citenamefont {Xu},
  \citenamefont {Bernhard}, \citenamefont {Mandrus}, \citenamefont {Qian},
  \citenamefont {Mele}, \citenamefont {Kane},\ and\ \citenamefont
  {Wu}}]{NiNatNanotech2021}%
  \BibitemOpen
  \bibfield  {author} {\bibinfo {author} {\bibfnamefont {Z.}~\bibnamefont
  {Ni}}, \bibinfo {author} {\bibfnamefont {A.~V.}\ \bibnamefont {Haglund}},
  \bibinfo {author} {\bibfnamefont {H.}~\bibnamefont {Wang}}, \bibinfo {author}
  {\bibfnamefont {B.}~\bibnamefont {Xu}}, \bibinfo {author} {\bibfnamefont
  {C.}~\bibnamefont {Bernhard}}, \bibinfo {author} {\bibfnamefont {D.~G.}\
  \bibnamefont {Mandrus}}, \bibinfo {author} {\bibfnamefont {X.}~\bibnamefont
  {Qian}}, \bibinfo {author} {\bibfnamefont {E.~J.}\ \bibnamefont {Mele}},
  \bibinfo {author} {\bibfnamefont {C.~L.}\ \bibnamefont {Kane}},\ and\
  \bibinfo {author} {\bibfnamefont {L.}~\bibnamefont {Wu}},\ }\href
  {https://doi.org/10.1038/s41565-021-00885-5} {\bibfield  {journal} {\bibinfo
  {journal} {Nature Nanotechnology}\ }\textbf {\bibinfo {volume} {16}},\
  \bibinfo {pages} {782} (\bibinfo {year} {2021})}\BibitemShut {NoStop}%
\bibitem [{\citenamefont {Kim}\ \emph {et~al.}(2019)\citenamefont {Kim},
  \citenamefont {Lim}, \citenamefont {Lee}, \citenamefont {Lee}, \citenamefont
  {Kim}, \citenamefont {Park}, \citenamefont {Jeon}, \citenamefont {Park},
  \citenamefont {Park},\ and\ \citenamefont {Cheong}}]{KimNatCommu2019}%
  \BibitemOpen
  \bibfield  {author} {\bibinfo {author} {\bibfnamefont {K.}~\bibnamefont
  {Kim}}, \bibinfo {author} {\bibfnamefont {S.~Y.}\ \bibnamefont {Lim}},
  \bibinfo {author} {\bibfnamefont {J.-U.}\ \bibnamefont {Lee}}, \bibinfo
  {author} {\bibfnamefont {S.}~\bibnamefont {Lee}}, \bibinfo {author}
  {\bibfnamefont {T.~Y.}\ \bibnamefont {Kim}}, \bibinfo {author} {\bibfnamefont
  {K.}~\bibnamefont {Park}}, \bibinfo {author} {\bibfnamefont {G.~S.}\
  \bibnamefont {Jeon}}, \bibinfo {author} {\bibfnamefont {C.-H.}\ \bibnamefont
  {Park}}, \bibinfo {author} {\bibfnamefont {J.-G.}\ \bibnamefont {Park}},\
  and\ \bibinfo {author} {\bibfnamefont {H.}~\bibnamefont {Cheong}},\ }\href
  {https://doi.org/10.1038/s41467-018-08284-6} {\bibfield  {journal} {\bibinfo
  {journal} {Nature Communications}\ }\textbf {\bibinfo {volume} {10}},\
  \bibinfo {pages} {345} (\bibinfo {year} {2019})}\BibitemShut {NoStop}%
\bibitem [{\citenamefont {Zhang}\ \emph {et~al.}(2022)\citenamefont {Zhang},
  \citenamefont {Ni}, \citenamefont {Stevens}, \citenamefont {Bai},
  \citenamefont {Peiris}, \citenamefont {Hendrickson}, \citenamefont {Wu},\
  and\ \citenamefont {Jariwala}}]{ZhangNatPhoto2022}%
  \BibitemOpen
  \bibfield  {author} {\bibinfo {author} {\bibfnamefont {H.}~\bibnamefont
  {Zhang}}, \bibinfo {author} {\bibfnamefont {Z.}~\bibnamefont {Ni}}, \bibinfo
  {author} {\bibfnamefont {C.~E.}\ \bibnamefont {Stevens}}, \bibinfo {author}
  {\bibfnamefont {A.}~\bibnamefont {Bai}}, \bibinfo {author} {\bibfnamefont
  {F.}~\bibnamefont {Peiris}}, \bibinfo {author} {\bibfnamefont {J.~R.}\
  \bibnamefont {Hendrickson}}, \bibinfo {author} {\bibfnamefont
  {L.}~\bibnamefont {Wu}},\ and\ \bibinfo {author} {\bibfnamefont
  {D.}~\bibnamefont {Jariwala}},\ }\href
  {https://doi.org/10.1038/s41566-022-00970-8} {\bibfield  {journal} {\bibinfo
  {journal} {Nature Photonics}\ }\textbf {\bibinfo {volume} {16}},\ \bibinfo
  {pages} {311} (\bibinfo {year} {2022})}\BibitemShut {NoStop}%
\bibitem [{\citenamefont {Zhang}\ \emph
  {et~al.}(2021{\natexlab{b}})\citenamefont {Zhang}, \citenamefont {Hwangbo},
  \citenamefont {Wang}, \citenamefont {Jiang}, \citenamefont {Chu},
  \citenamefont {Wen}, \citenamefont {Xiao},\ and\ \citenamefont
  {Xu}}]{ZhangNanoLett2021}%
  \BibitemOpen
  \bibfield  {author} {\bibinfo {author} {\bibfnamefont {Q.}~\bibnamefont
  {Zhang}}, \bibinfo {author} {\bibfnamefont {K.}~\bibnamefont {Hwangbo}},
  \bibinfo {author} {\bibfnamefont {C.}~\bibnamefont {Wang}}, \bibinfo {author}
  {\bibfnamefont {Q.}~\bibnamefont {Jiang}}, \bibinfo {author} {\bibfnamefont
  {J.-H.}\ \bibnamefont {Chu}}, \bibinfo {author} {\bibfnamefont
  {H.}~\bibnamefont {Wen}}, \bibinfo {author} {\bibfnamefont {D.}~\bibnamefont
  {Xiao}},\ and\ \bibinfo {author} {\bibfnamefont {X.}~\bibnamefont {Xu}},\
  }\href {https://doi.org/10.1021/acs.nanolett.1c02188} {\bibfield  {journal}
  {\bibinfo  {journal} {Nano Letters}\ }\textbf {\bibinfo {volume} {21}},\
  \bibinfo {pages} {6938} (\bibinfo {year} {2021}{\natexlab{b}})}\BibitemShut
  {NoStop}%
\bibitem [{NiP()}]{NiPRL2021}%
  \BibitemOpen
  \href@noop {} {\ }\BibitemShut {NoStop}%
\bibitem [{\citenamefont {Chu}\ \emph {et~al.}(2020)\citenamefont {Chu},
  \citenamefont {Roh}, \citenamefont {Island}, \citenamefont {Li},
  \citenamefont {Lee}, \citenamefont {Chen}, \citenamefont {Park},
  \citenamefont {Young}, \citenamefont {Lee},\ and\ \citenamefont
  {Hsieh}}]{ChuPRL2020}%
  \BibitemOpen
  \bibfield  {author} {\bibinfo {author} {\bibfnamefont {H.}~\bibnamefont
  {Chu}}, \bibinfo {author} {\bibfnamefont {C.~J.}\ \bibnamefont {Roh}},
  \bibinfo {author} {\bibfnamefont {J.~O.}\ \bibnamefont {Island}}, \bibinfo
  {author} {\bibfnamefont {C.}~\bibnamefont {Li}}, \bibinfo {author}
  {\bibfnamefont {S.}~\bibnamefont {Lee}}, \bibinfo {author} {\bibfnamefont
  {J.}~\bibnamefont {Chen}}, \bibinfo {author} {\bibfnamefont {J.-G.}\
  \bibnamefont {Park}}, \bibinfo {author} {\bibfnamefont {A.~F.}\ \bibnamefont
  {Young}}, \bibinfo {author} {\bibfnamefont {J.~S.}\ \bibnamefont {Lee}},\
  and\ \bibinfo {author} {\bibfnamefont {D.}~\bibnamefont {Hsieh}},\ }\href
  {https://doi.org/10.1103/PhysRevLett.124.027601} {\bibfield  {journal}
  {\bibinfo  {journal} {Phys. Rev. Lett.}\ }\textbf {\bibinfo {volume} {124}},\
  \bibinfo {pages} {027601} (\bibinfo {year} {2020})}\BibitemShut {NoStop}%
\bibitem [{\citenamefont {Shan}\ \emph {et~al.}(2021)\citenamefont {Shan},
  \citenamefont {Ye}, \citenamefont {Chu}, \citenamefont {Lee}, \citenamefont
  {Park}, \citenamefont {Balents},\ and\ \citenamefont
  {Hsieh}}]{ShanNature2021}%
  \BibitemOpen
  \bibfield  {author} {\bibinfo {author} {\bibfnamefont {J.-Y.}\ \bibnamefont
  {Shan}}, \bibinfo {author} {\bibfnamefont {M.}~\bibnamefont {Ye}}, \bibinfo
  {author} {\bibfnamefont {H.}~\bibnamefont {Chu}}, \bibinfo {author}
  {\bibfnamefont {S.}~\bibnamefont {Lee}}, \bibinfo {author} {\bibfnamefont
  {J.-G.}\ \bibnamefont {Park}}, \bibinfo {author} {\bibfnamefont
  {L.}~\bibnamefont {Balents}},\ and\ \bibinfo {author} {\bibfnamefont
  {D.}~\bibnamefont {Hsieh}},\ }\href
  {https://doi.org/10.1038/s41586-021-04051-8} {\bibfield  {journal} {\bibinfo
  {journal} {Nature}\ }\textbf {\bibinfo {volume} {600}},\ \bibinfo {pages}
  {235} (\bibinfo {year} {2021})}\BibitemShut {NoStop}%
\bibitem [{\citenamefont {Wang}\ \emph {et~al.}(2023)\citenamefont {Wang},
  \citenamefont {Zhang}, \citenamefont {Shiomi}, \citenamefont {hisa Arima},
  \citenamefont {Nagaosa}, \citenamefont {Tokura},\ and\ \citenamefont
  {Ogawa}}]{WangArxiv2023}%
  \BibitemOpen
  \bibfield  {author} {\bibinfo {author} {\bibfnamefont {Z.}~\bibnamefont
  {Wang}}, \bibinfo {author} {\bibfnamefont {X.-X.}\ \bibnamefont {Zhang}},
  \bibinfo {author} {\bibfnamefont {Y.}~\bibnamefont {Shiomi}}, \bibinfo
  {author} {\bibfnamefont {T.}~\bibnamefont {hisa Arima}}, \bibinfo {author}
  {\bibfnamefont {N.}~\bibnamefont {Nagaosa}}, \bibinfo {author} {\bibfnamefont
  {Y.}~\bibnamefont {Tokura}},\ and\ \bibinfo {author} {\bibfnamefont
  {N.}~\bibnamefont {Ogawa}},\ }\href@noop {} {\bibinfo {title} {Exciton-magnon
  splitting in van der waals antiferromagnet $\mathrm{MnPS_3}$ unveiled by
  second-harmonic generation}} (\bibinfo {year} {2023}),\ \Eprint
  {https://arxiv.org/abs/2306.14642} {arXiv:2306.14642 [cond-mat.str-el]}
  \BibitemShut {NoStop}%
\bibitem [{\citenamefont {Yoshikawa}\ \emph {et~al.}(2017)\citenamefont
  {Yoshikawa}, \citenamefont {Tamaya},\ and\ \citenamefont
  {Tanaka}}]{NaotakaScience2017}%
  \BibitemOpen
  \bibfield  {author} {\bibinfo {author} {\bibfnamefont {N.}~\bibnamefont
  {Yoshikawa}}, \bibinfo {author} {\bibfnamefont {T.}~\bibnamefont {Tamaya}},\
  and\ \bibinfo {author} {\bibfnamefont {K.}~\bibnamefont {Tanaka}},\ }\href
  {https://doi.org/10.1126/science.aam8861} {\bibfield  {journal} {\bibinfo
  {journal} {Science}\ }\textbf {\bibinfo {volume} {356}},\ \bibinfo {pages}
  {736} (\bibinfo {year} {2017})}\BibitemShut {NoStop}%
\bibitem [{\citenamefont {Liu}\ \emph {et~al.}(2017)\citenamefont {Liu},
  \citenamefont {Li}, \citenamefont {You}, \citenamefont {Ghimire},
  \citenamefont {Heinz},\ and\ \citenamefont {Reis}}]{LiuNatPhys2017}%
  \BibitemOpen
  \bibfield  {author} {\bibinfo {author} {\bibfnamefont {H.}~\bibnamefont
  {Liu}}, \bibinfo {author} {\bibfnamefont {Y.}~\bibnamefont {Li}}, \bibinfo
  {author} {\bibfnamefont {Y.~S.}\ \bibnamefont {You}}, \bibinfo {author}
  {\bibfnamefont {S.}~\bibnamefont {Ghimire}}, \bibinfo {author} {\bibfnamefont
  {T.~F.}\ \bibnamefont {Heinz}},\ and\ \bibinfo {author} {\bibfnamefont
  {D.~A.}\ \bibnamefont {Reis}},\ }\href {https://doi.org/10.1038/nphys3946}
  {\bibfield  {journal} {\bibinfo  {journal} {Nature Physics}\ }\textbf
  {\bibinfo {volume} {13}},\ \bibinfo {pages} {262} (\bibinfo {year}
  {2017})}\BibitemShut {NoStop}%
\bibitem [{\citenamefont {Hong}\ \emph {et~al.}(2023)\citenamefont {Hong},
  \citenamefont {Huang}, \citenamefont {Ma}, \citenamefont {Qi}, \citenamefont
  {Liu}, \citenamefont {Wu}, \citenamefont {Sun}, \citenamefont {Wang},\ and\
  \citenamefont {Liu}}]{HongArxiv2023}%
  \BibitemOpen
  \bibfield  {author} {\bibinfo {author} {\bibfnamefont {H.}~\bibnamefont
  {Hong}}, \bibinfo {author} {\bibfnamefont {C.}~\bibnamefont {Huang}},
  \bibinfo {author} {\bibfnamefont {C.}~\bibnamefont {Ma}}, \bibinfo {author}
  {\bibfnamefont {J.}~\bibnamefont {Qi}}, \bibinfo {author} {\bibfnamefont
  {C.}~\bibnamefont {Liu}}, \bibinfo {author} {\bibfnamefont {S.}~\bibnamefont
  {Wu}}, \bibinfo {author} {\bibfnamefont {Z.}~\bibnamefont {Sun}}, \bibinfo
  {author} {\bibfnamefont {E.}~\bibnamefont {Wang}},\ and\ \bibinfo {author}
  {\bibfnamefont {K.}~\bibnamefont {Liu}},\ }\href@noop {} {\bibinfo {title}
  {Twist-phase-matching in two-dimensional materials}} (\bibinfo {year}
  {2023}),\ \Eprint {https://arxiv.org/abs/2305.11511} {arXiv:2305.11511
  [physics.optics]} \BibitemShut {NoStop}%
\bibitem [{\citenamefont {Dudley}\ \emph {et~al.}(2010)\citenamefont {Dudley},
  \citenamefont {Genty},\ and\ \citenamefont {St\'ephane}}]{DudleyBook}%
  \BibitemOpen
  \bibfield  {author} {\bibinfo {author} {\bibfnamefont {J.~M.}\ \bibnamefont
  {Dudley}}, \bibinfo {author} {\bibfnamefont {G.}~\bibnamefont {Genty}},\ and\
  \bibinfo {author} {\bibfnamefont {C.}~\bibnamefont {St\'ephane}},\
  }\href@noop {} {\emph {\bibinfo {title} {Supercontinuum Generation in Optical
  Fibers}}}\ (\bibinfo  {publisher} {Cambridge U. Press, New York},\ \bibinfo
  {year} {2010})\BibitemShut {NoStop}%
\bibitem [{\citenamefont {Dudley}\ \emph {et~al.}(2006)\citenamefont {Dudley},
  \citenamefont {Genty},\ and\ \citenamefont {Coen}}]{DudleyRevModPhys2006}%
  \BibitemOpen
  \bibfield  {author} {\bibinfo {author} {\bibfnamefont {J.~M.}\ \bibnamefont
  {Dudley}}, \bibinfo {author} {\bibfnamefont {G.}~\bibnamefont {Genty}},\ and\
  \bibinfo {author} {\bibfnamefont {S.}~\bibnamefont {Coen}},\ }\href
  {https://doi.org/10.1103/RevModPhys.78.1135} {\bibfield  {journal} {\bibinfo
  {journal} {Rev. Mod. Phys.}\ }\textbf {\bibinfo {volume} {78}},\ \bibinfo
  {pages} {1135} (\bibinfo {year} {2006})}\BibitemShut {NoStop}%
\bibitem [{\citenamefont {Dudley}\ and\ \citenamefont
  {Genty}(2013)}]{DudleyPhysToday2013}%
  \BibitemOpen
  \bibfield  {author} {\bibinfo {author} {\bibfnamefont {J.~M.}\ \bibnamefont
  {Dudley}}\ and\ \bibinfo {author} {\bibfnamefont {G.}~\bibnamefont {Genty}},\
  }\href {https://doi.org/10.1063/PT.3.2045} {\bibfield  {journal} {\bibinfo
  {journal} {Physics Today}\ }\textbf {\bibinfo {volume} {66}},\ \bibinfo
  {pages} {29} (\bibinfo {year} {2013})}\BibitemShut {NoStop}%
\bibitem [{\citenamefont {Chittari}\ \emph {et~al.}(2016)\citenamefont
  {Chittari}, \citenamefont {Park}, \citenamefont {Lee}, \citenamefont {Han},
  \citenamefont {MacDonald}, \citenamefont {Hwang},\ and\ \citenamefont
  {Jung}}]{ChittariPRB2016}%
  \BibitemOpen
  \bibfield  {author} {\bibinfo {author} {\bibfnamefont {B.~L.}\ \bibnamefont
  {Chittari}}, \bibinfo {author} {\bibfnamefont {Y.}~\bibnamefont {Park}},
  \bibinfo {author} {\bibfnamefont {D.}~\bibnamefont {Lee}}, \bibinfo {author}
  {\bibfnamefont {M.}~\bibnamefont {Han}}, \bibinfo {author} {\bibfnamefont
  {A.~H.}\ \bibnamefont {MacDonald}}, \bibinfo {author} {\bibfnamefont
  {E.}~\bibnamefont {Hwang}},\ and\ \bibinfo {author} {\bibfnamefont
  {J.}~\bibnamefont {Jung}},\ }\href
  {https://doi.org/10.1103/PhysRevB.94.184428} {\bibfield  {journal} {\bibinfo
  {journal} {Phys. Rev. B}\ }\textbf {\bibinfo {volume} {94}},\ \bibinfo
  {pages} {184428} (\bibinfo {year} {2016})}\BibitemShut {NoStop}%
\bibitem [{\citenamefont {Huang}\ \emph {et~al.}(2022)\citenamefont {Huang},
  \citenamefont {Choi}, \citenamefont {Shih},\ and\ \citenamefont
  {Li}}]{HuangNatNanotech2022}%
  \BibitemOpen
  \bibfield  {author} {\bibinfo {author} {\bibfnamefont {D.}~\bibnamefont
  {Huang}}, \bibinfo {author} {\bibfnamefont {J.}~\bibnamefont {Choi}},
  \bibinfo {author} {\bibfnamefont {C.-K.}\ \bibnamefont {Shih}},\ and\
  \bibinfo {author} {\bibfnamefont {X.}~\bibnamefont {Li}},\ }\href
  {https://doi.org/10.1038/s41565-021-01068-y} {\bibfield  {journal} {\bibinfo
  {journal} {Nature Nanotechnology}\ }\textbf {\bibinfo {volume} {17}},\
  \bibinfo {pages} {227} (\bibinfo {year} {2022})}\BibitemShut {NoStop}%
\bibitem [{\citenamefont {Xu}\ \emph {et~al.}(2022)\citenamefont {Xu},
  \citenamefont {Trovatello}, \citenamefont {Mooshammer}, \citenamefont {Shao},
  \citenamefont {Zhang}, \citenamefont {Yao}, \citenamefont {Basov},
  \citenamefont {Cerullo},\ and\ \citenamefont {Schuck}}]{XuNatNano2022}%
  \BibitemOpen
  \bibfield  {author} {\bibinfo {author} {\bibfnamefont {X.}~\bibnamefont
  {Xu}}, \bibinfo {author} {\bibfnamefont {C.}~\bibnamefont {Trovatello}},
  \bibinfo {author} {\bibfnamefont {F.}~\bibnamefont {Mooshammer}}, \bibinfo
  {author} {\bibfnamefont {Y.}~\bibnamefont {Shao}}, \bibinfo {author}
  {\bibfnamefont {S.}~\bibnamefont {Zhang}}, \bibinfo {author} {\bibfnamefont
  {K.}~\bibnamefont {Yao}}, \bibinfo {author} {\bibfnamefont {D.~N.}\
  \bibnamefont {Basov}}, \bibinfo {author} {\bibfnamefont {G.}~\bibnamefont
  {Cerullo}},\ and\ \bibinfo {author} {\bibfnamefont {P.~J.}\ \bibnamefont
  {Schuck}},\ }\href {https://doi.org/10.1038/s41566-022-01053-4} {\bibfield
  {journal} {\bibinfo  {journal} {Nature Photonics}\ }\textbf {\bibinfo
  {volume} {16}},\ \bibinfo {pages} {698} (\bibinfo {year} {2022})}\BibitemShut
  {NoStop}%
\bibitem [{\citenamefont {Sun}\ \emph {et~al.}(2019)\citenamefont {Sun},
  \citenamefont {Yi}, \citenamefont {Song}, \citenamefont {Clark},
  \citenamefont {Huang}, \citenamefont {Shan}, \citenamefont {Wu},
  \citenamefont {Huang}, \citenamefont {Gao}, \citenamefont {Chen},
  \citenamefont {McGuire}, \citenamefont {Cao}, \citenamefont {Xiao},
  \citenamefont {Liu}, \citenamefont {Yao}, \citenamefont {Xu},\ and\
  \citenamefont {Wu}}]{SunNature2019}%
  \BibitemOpen
  \bibfield  {author} {\bibinfo {author} {\bibfnamefont {Z.}~\bibnamefont
  {Sun}}, \bibinfo {author} {\bibfnamefont {Y.}~\bibnamefont {Yi}}, \bibinfo
  {author} {\bibfnamefont {T.}~\bibnamefont {Song}}, \bibinfo {author}
  {\bibfnamefont {G.}~\bibnamefont {Clark}}, \bibinfo {author} {\bibfnamefont
  {B.}~\bibnamefont {Huang}}, \bibinfo {author} {\bibfnamefont
  {Y.}~\bibnamefont {Shan}}, \bibinfo {author} {\bibfnamefont {S.}~\bibnamefont
  {Wu}}, \bibinfo {author} {\bibfnamefont {D.}~\bibnamefont {Huang}}, \bibinfo
  {author} {\bibfnamefont {C.}~\bibnamefont {Gao}}, \bibinfo {author}
  {\bibfnamefont {Z.}~\bibnamefont {Chen}}, \bibinfo {author} {\bibfnamefont
  {M.}~\bibnamefont {McGuire}}, \bibinfo {author} {\bibfnamefont
  {T.}~\bibnamefont {Cao}}, \bibinfo {author} {\bibfnamefont {D.}~\bibnamefont
  {Xiao}}, \bibinfo {author} {\bibfnamefont {W.-T.}\ \bibnamefont {Liu}},
  \bibinfo {author} {\bibfnamefont {W.}~\bibnamefont {Yao}}, \bibinfo {author}
  {\bibfnamefont {X.}~\bibnamefont {Xu}},\ and\ \bibinfo {author}
  {\bibfnamefont {S.}~\bibnamefont {Wu}},\ }\href
  {https://doi.org/10.1038/s41586-019-1445-3} {\bibfield  {journal} {\bibinfo
  {journal} {Nature}\ }\textbf {\bibinfo {volume} {572}},\ \bibinfo {pages}
  {497} (\bibinfo {year} {2019})}\BibitemShut {NoStop}%
\bibitem [{\citenamefont {Low}\ \emph {et~al.}(2017)\citenamefont {Low},
  \citenamefont {Chaves}, \citenamefont {Caldwell}, \citenamefont {Kumar},
  \citenamefont {Fang}, \citenamefont {Avouris}, \citenamefont {Heinz},
  \citenamefont {Guinea}, \citenamefont {Martin-Moreno},\ and\ \citenamefont
  {Koppens}}]{LowNatMater2017}%
  \BibitemOpen
  \bibfield  {author} {\bibinfo {author} {\bibfnamefont {T.}~\bibnamefont
  {Low}}, \bibinfo {author} {\bibfnamefont {A.}~\bibnamefont {Chaves}},
  \bibinfo {author} {\bibfnamefont {J.~D.}\ \bibnamefont {Caldwell}}, \bibinfo
  {author} {\bibfnamefont {A.}~\bibnamefont {Kumar}}, \bibinfo {author}
  {\bibfnamefont {N.~X.}\ \bibnamefont {Fang}}, \bibinfo {author}
  {\bibfnamefont {P.}~\bibnamefont {Avouris}}, \bibinfo {author} {\bibfnamefont
  {T.~F.}\ \bibnamefont {Heinz}}, \bibinfo {author} {\bibfnamefont
  {F.}~\bibnamefont {Guinea}}, \bibinfo {author} {\bibfnamefont
  {L.}~\bibnamefont {Martin-Moreno}},\ and\ \bibinfo {author} {\bibfnamefont
  {F.}~\bibnamefont {Koppens}},\ }\href {https://doi.org/10.1038/nmat4792}
  {\bibfield  {journal} {\bibinfo  {journal} {Nature Materials}\ }\textbf
  {\bibinfo {volume} {16}},\ \bibinfo {pages} {182} (\bibinfo {year}
  {2017})}\BibitemShut {NoStop}%
\bibitem [{\citenamefont {Regan}\ \emph {et~al.}(2022)\citenamefont {Regan},
  \citenamefont {Wang}, \citenamefont {Paik}, \citenamefont {Zeng},
  \citenamefont {Zhang}, \citenamefont {Zhu}, \citenamefont {MacDonald},
  \citenamefont {Deng},\ and\ \citenamefont {Wang}}]{ReganNatRevMater2022}%
  \BibitemOpen
  \bibfield  {author} {\bibinfo {author} {\bibfnamefont {E.~C.}\ \bibnamefont
  {Regan}}, \bibinfo {author} {\bibfnamefont {D.}~\bibnamefont {Wang}},
  \bibinfo {author} {\bibfnamefont {E.~Y.}\ \bibnamefont {Paik}}, \bibinfo
  {author} {\bibfnamefont {Y.}~\bibnamefont {Zeng}}, \bibinfo {author}
  {\bibfnamefont {L.}~\bibnamefont {Zhang}}, \bibinfo {author} {\bibfnamefont
  {J.}~\bibnamefont {Zhu}}, \bibinfo {author} {\bibfnamefont {A.~H.}\
  \bibnamefont {MacDonald}}, \bibinfo {author} {\bibfnamefont {H.}~\bibnamefont
  {Deng}},\ and\ \bibinfo {author} {\bibfnamefont {F.}~\bibnamefont {Wang}},\
  }\href {https://doi.org/10.1038/s41578-022-00440-1} {\bibfield  {journal}
  {\bibinfo  {journal} {Nature Reviews Materials}\ }\textbf {\bibinfo {volume}
  {7}},\ \bibinfo {pages} {778} (\bibinfo {year} {2022})}\BibitemShut {NoStop}%
\bibitem [{\citenamefont {Langford}\ \emph {et~al.}(2011)\citenamefont
  {Langford}, \citenamefont {Ramelow}, \citenamefont {Prevedel}, \citenamefont
  {Munro}, \citenamefont {Milburn},\ and\ \citenamefont
  {Zeilinger}}]{LangfordNature2011}%
  \BibitemOpen
  \bibfield  {author} {\bibinfo {author} {\bibfnamefont {N.~K.}\ \bibnamefont
  {Langford}}, \bibinfo {author} {\bibfnamefont {S.}~\bibnamefont {Ramelow}},
  \bibinfo {author} {\bibfnamefont {R.}~\bibnamefont {Prevedel}}, \bibinfo
  {author} {\bibfnamefont {W.~J.}\ \bibnamefont {Munro}}, \bibinfo {author}
  {\bibfnamefont {G.~J.}\ \bibnamefont {Milburn}},\ and\ \bibinfo {author}
  {\bibfnamefont {A.}~\bibnamefont {Zeilinger}},\ }\href
  {https://doi.org/10.1038/nature10463} {\bibfield  {journal} {\bibinfo
  {journal} {Nature}\ }\textbf {\bibinfo {volume} {478}},\ \bibinfo {pages}
  {360} (\bibinfo {year} {2011})}\BibitemShut {NoStop}%
\bibitem [{\citenamefont {Costello}\ \emph {et~al.}(2021)\citenamefont
  {Costello}, \citenamefont {O'Hara}, \citenamefont {Wu}, \citenamefont
  {Valovcin}, \citenamefont {Pfeiffer}, \citenamefont {West},\ and\
  \citenamefont {Sherwin}}]{CostelloNature2021}%
  \BibitemOpen
  \bibfield  {author} {\bibinfo {author} {\bibfnamefont {J.~B.}\ \bibnamefont
  {Costello}}, \bibinfo {author} {\bibfnamefont {S.~D.}\ \bibnamefont
  {O'Hara}}, \bibinfo {author} {\bibfnamefont {Q.}~\bibnamefont {Wu}}, \bibinfo
  {author} {\bibfnamefont {D.~C.}\ \bibnamefont {Valovcin}}, \bibinfo {author}
  {\bibfnamefont {L.~N.}\ \bibnamefont {Pfeiffer}}, \bibinfo {author}
  {\bibfnamefont {K.~W.}\ \bibnamefont {West}},\ and\ \bibinfo {author}
  {\bibfnamefont {M.~S.}\ \bibnamefont {Sherwin}},\ }\href
  {https://doi.org/10.1038/s41586-021-03940-2} {\bibfield  {journal} {\bibinfo
  {journal} {Nature}\ }\textbf {\bibinfo {volume} {599}},\ \bibinfo {pages}
  {57} (\bibinfo {year} {2021})}\BibitemShut {NoStop}%
\bibitem [{\citenamefont {Yue}\ \emph {et~al.}(2022)\citenamefont {Yue},
  \citenamefont {Hollinger}, \citenamefont {Uzundal}, \citenamefont {Nebgen},
  \citenamefont {Gan}, \citenamefont {Najafidehaghani}, \citenamefont {George},
  \citenamefont {Spielmann}, \citenamefont {Kartashov}, \citenamefont
  {Turchanin}, \citenamefont {Qiu}, \citenamefont {Gaarde},\ and\ \citenamefont
  {Zuerch}}]{YuePRL2022}%
  \BibitemOpen
  \bibfield  {author} {\bibinfo {author} {\bibfnamefont {L.}~\bibnamefont
  {Yue}}, \bibinfo {author} {\bibfnamefont {R.}~\bibnamefont {Hollinger}},
  \bibinfo {author} {\bibfnamefont {C.~B.}\ \bibnamefont {Uzundal}}, \bibinfo
  {author} {\bibfnamefont {B.}~\bibnamefont {Nebgen}}, \bibinfo {author}
  {\bibfnamefont {Z.}~\bibnamefont {Gan}}, \bibinfo {author} {\bibfnamefont
  {E.}~\bibnamefont {Najafidehaghani}}, \bibinfo {author} {\bibfnamefont
  {A.}~\bibnamefont {George}}, \bibinfo {author} {\bibfnamefont
  {C.}~\bibnamefont {Spielmann}}, \bibinfo {author} {\bibfnamefont
  {D.}~\bibnamefont {Kartashov}}, \bibinfo {author} {\bibfnamefont
  {A.}~\bibnamefont {Turchanin}}, \bibinfo {author} {\bibfnamefont {D.~Y.}\
  \bibnamefont {Qiu}}, \bibinfo {author} {\bibfnamefont {M.~B.}\ \bibnamefont
  {Gaarde}},\ and\ \bibinfo {author} {\bibfnamefont {M.}~\bibnamefont
  {Zuerch}},\ }\href {https://doi.org/10.1103/PhysRevLett.129.147401}
  {\bibfield  {journal} {\bibinfo  {journal} {Phys. Rev. Lett.}\ }\textbf
  {\bibinfo {volume} {129}},\ \bibinfo {pages} {147401} (\bibinfo {year}
  {2022})}\BibitemShut {NoStop}%
\bibitem [{\citenamefont {Schubert}\ \emph {et~al.}(2014)\citenamefont
  {Schubert}, \citenamefont {Hohenleutner}, \citenamefont {Langer},
  \citenamefont {Urbanek}, \citenamefont {Lange}, \citenamefont {Huttner},
  \citenamefont {Golde}, \citenamefont {Meier}, \citenamefont {Kira},
  \citenamefont {Koch},\ and\ \citenamefont {Huber}}]{SchubertNatPhoto2014}%
  \BibitemOpen
  \bibfield  {author} {\bibinfo {author} {\bibfnamefont {O.}~\bibnamefont
  {Schubert}}, \bibinfo {author} {\bibfnamefont {M.}~\bibnamefont
  {Hohenleutner}}, \bibinfo {author} {\bibfnamefont {F.}~\bibnamefont
  {Langer}}, \bibinfo {author} {\bibfnamefont {B.}~\bibnamefont {Urbanek}},
  \bibinfo {author} {\bibfnamefont {C.}~\bibnamefont {Lange}}, \bibinfo
  {author} {\bibfnamefont {U.}~\bibnamefont {Huttner}}, \bibinfo {author}
  {\bibfnamefont {D.}~\bibnamefont {Golde}}, \bibinfo {author} {\bibfnamefont
  {T.}~\bibnamefont {Meier}}, \bibinfo {author} {\bibfnamefont
  {M.}~\bibnamefont {Kira}}, \bibinfo {author} {\bibfnamefont {S.~W.}\
  \bibnamefont {Koch}},\ and\ \bibinfo {author} {\bibfnamefont
  {R.}~\bibnamefont {Huber}},\ }\href
  {https://doi.org/10.1038/nphoton.2013.349} {\bibfield  {journal} {\bibinfo
  {journal} {Nature Photonics}\ }\textbf {\bibinfo {volume} {8}},\ \bibinfo
  {pages} {119} (\bibinfo {year} {2014})}\BibitemShut {NoStop}%
\bibitem [{\citenamefont {Langer}\ \emph {et~al.}(2016)\citenamefont {Langer},
  \citenamefont {Hohenleutner}, \citenamefont {Schmid}, \citenamefont
  {Poellmann}, \citenamefont {Nagler}, \citenamefont {Korn}, \citenamefont
  {Sch{\"u}ller}, \citenamefont {Sherwin}, \citenamefont {Huttner},
  \citenamefont {Steiner}, \citenamefont {Koch}, \citenamefont {Kira},\ and\
  \citenamefont {Huber}}]{LangerNature2016}%
  \BibitemOpen
  \bibfield  {author} {\bibinfo {author} {\bibfnamefont {F.}~\bibnamefont
  {Langer}}, \bibinfo {author} {\bibfnamefont {M.}~\bibnamefont
  {Hohenleutner}}, \bibinfo {author} {\bibfnamefont {C.~P.}\ \bibnamefont
  {Schmid}}, \bibinfo {author} {\bibfnamefont {C.}~\bibnamefont {Poellmann}},
  \bibinfo {author} {\bibfnamefont {P.}~\bibnamefont {Nagler}}, \bibinfo
  {author} {\bibfnamefont {T.}~\bibnamefont {Korn}}, \bibinfo {author}
  {\bibfnamefont {C.}~\bibnamefont {Sch{\"u}ller}}, \bibinfo {author}
  {\bibfnamefont {M.~S.}\ \bibnamefont {Sherwin}}, \bibinfo {author}
  {\bibfnamefont {U.}~\bibnamefont {Huttner}}, \bibinfo {author} {\bibfnamefont
  {J.~T.}\ \bibnamefont {Steiner}}, \bibinfo {author} {\bibfnamefont {S.~W.}\
  \bibnamefont {Koch}}, \bibinfo {author} {\bibfnamefont {M.}~\bibnamefont
  {Kira}},\ and\ \bibinfo {author} {\bibfnamefont {R.}~\bibnamefont {Huber}},\
  }\href {https://doi.org/10.1038/nature17958} {\bibfield  {journal} {\bibinfo
  {journal} {Nature}\ }\textbf {\bibinfo {volume} {533}},\ \bibinfo {pages}
  {225} (\bibinfo {year} {2016})}\BibitemShut {NoStop}%
\bibitem [{\citenamefont {Langer}\ \emph {et~al.}(2018)\citenamefont {Langer},
  \citenamefont {Schmid}, \citenamefont {Schlauderer}, \citenamefont {Gmitra},
  \citenamefont {Fabian}, \citenamefont {Nagler}, \citenamefont {Sch{\"u}ller},
  \citenamefont {Korn}, \citenamefont {Hawkins}, \citenamefont {Steiner},
  \citenamefont {Huttner}, \citenamefont {Koch}, \citenamefont {Kira},\ and\
  \citenamefont {Huber}}]{LangerNature2018}%
  \BibitemOpen
  \bibfield  {author} {\bibinfo {author} {\bibfnamefont {F.}~\bibnamefont
  {Langer}}, \bibinfo {author} {\bibfnamefont {C.~P.}\ \bibnamefont {Schmid}},
  \bibinfo {author} {\bibfnamefont {S.}~\bibnamefont {Schlauderer}}, \bibinfo
  {author} {\bibfnamefont {M.}~\bibnamefont {Gmitra}}, \bibinfo {author}
  {\bibfnamefont {J.}~\bibnamefont {Fabian}}, \bibinfo {author} {\bibfnamefont
  {P.}~\bibnamefont {Nagler}}, \bibinfo {author} {\bibfnamefont
  {C.}~\bibnamefont {Sch{\"u}ller}}, \bibinfo {author} {\bibfnamefont
  {T.}~\bibnamefont {Korn}}, \bibinfo {author} {\bibfnamefont {P.~G.}\
  \bibnamefont {Hawkins}}, \bibinfo {author} {\bibfnamefont {J.~T.}\
  \bibnamefont {Steiner}}, \bibinfo {author} {\bibfnamefont {U.}~\bibnamefont
  {Huttner}}, \bibinfo {author} {\bibfnamefont {S.~W.}\ \bibnamefont {Koch}},
  \bibinfo {author} {\bibfnamefont {M.}~\bibnamefont {Kira}},\ and\ \bibinfo
  {author} {\bibfnamefont {R.}~\bibnamefont {Huber}},\ }\href
  {https://doi.org/10.1038/s41586-018-0013-6} {\bibfield  {journal} {\bibinfo
  {journal} {Nature}\ }\textbf {\bibinfo {volume} {557}},\ \bibinfo {pages}
  {76} (\bibinfo {year} {2018})}\BibitemShut {NoStop}%
\bibitem [{\citenamefont {Uchida}\ \emph {et~al.}(2016)\citenamefont {Uchida},
  \citenamefont {Otobe}, \citenamefont {Mochizuki}, \citenamefont {Kim},
  \citenamefont {Yoshita}, \citenamefont {Akiyama}, \citenamefont {Pfeiffer},
  \citenamefont {West}, \citenamefont {Tanaka},\ and\ \citenamefont
  {Hirori}}]{UchidaPRL2016}%
  \BibitemOpen
  \bibfield  {author} {\bibinfo {author} {\bibfnamefont {K.}~\bibnamefont
  {Uchida}}, \bibinfo {author} {\bibfnamefont {T.}~\bibnamefont {Otobe}},
  \bibinfo {author} {\bibfnamefont {T.}~\bibnamefont {Mochizuki}}, \bibinfo
  {author} {\bibfnamefont {C.}~\bibnamefont {Kim}}, \bibinfo {author}
  {\bibfnamefont {M.}~\bibnamefont {Yoshita}}, \bibinfo {author} {\bibfnamefont
  {H.}~\bibnamefont {Akiyama}}, \bibinfo {author} {\bibfnamefont {L.~N.}\
  \bibnamefont {Pfeiffer}}, \bibinfo {author} {\bibfnamefont {K.~W.}\
  \bibnamefont {West}}, \bibinfo {author} {\bibfnamefont {K.}~\bibnamefont
  {Tanaka}},\ and\ \bibinfo {author} {\bibfnamefont {H.}~\bibnamefont
  {Hirori}},\ }\href {https://doi.org/10.1103/PhysRevLett.117.277402}
  {\bibfield  {journal} {\bibinfo  {journal} {Phys. Rev. Lett.}\ }\textbf
  {\bibinfo {volume} {117}},\ \bibinfo {pages} {277402} (\bibinfo {year}
  {2016})}\BibitemShut {NoStop}%
\end{thebibliography}%
\clearpage
\section*{Methods}
\textbf{Sample preparation.} Single crystals of MnPSe$_3$ and MnPS$_3$ were prepared by a chemical vapor transport method. A mixture of Mn (99.9\%), P (99\%), Se/S (99.9\%) with the molar ratios of Mn : P : Se/S = 1 : 1 : 3, and additional iodine as transport agent. The mixture was subsequently sealed in an evacuated quartz tube, which was placed in a two-zone furnace. The reaction zone was heated to 973 K and held for 5 days, with the growth zone held at 1023 K. Then exchange temperature gradient of the two zones and kept for 3 weeks. Then the quartz tube was naturally cooled down to room temperature. Finally, MnPSe$_3$ and MnPS$_3$ platelet-shaped crystals in several millimeter sizes were obtained in the growth zone.

The thin MnPSe$_3$ films (film1 and film2 in Fig.~3 ) were obtained by mechanical exfoliation on fused silica substrates. Film2 contains areas with different thickness ($\sim 80$ nm, $\sim 90$ nm, $\sim 100$ nm, see Supplementary Fig.~5), with the largest area of 100 nm thickness. In this study, there are other materials used as comparisons to MnPSe$_3$ samples. LiNbO$_3$ and fused silica substrates were commercially purchased. The 3R-WS$_2$ films were grown by the chemical vapor deposition method. The 80 nm GaSe film was obtained by mechanical exfoliation on fused silica substrate.

\textbf{Nonlinear optical wave mixing.} The 847 nm and 1280 nm lasers were generated by a Light-Conversion ORPHEUS-F hybrid optical parametric amplifier, pumped by 1024 nm laser generated by a Light Conversion PHAROS Yb:KGW laser system. The repetition rate was 50 kHz. The two lasers were focused onto the sample in a non-collinear geometry by two optical lens. The diameter of the laser spot at the focus was about 100 $\mu$m. The time resolution of the setup was $\sim 43$ fs. The two incident lasers were linearly polarized along $s$ direction.
The spectra of wave mixing signals were measured by a NOVA high sensitive spectrometer, ideaoptics, China. The spectrometer covers the 200-980 nm range. When measuring signals below 650 nm, short-pass color-glass filters were placed before the spectrometer to eliminate the contamination of spectrum by the scattered 847 nm excitation laser. The short-pass filters do not extend to infinitely short wavelengths but show cutoff near 300 nm. Therefore the overall accessible wavelength range of the measurement setup was 300-980 nm. See Supplementary Fig.~1 for the schematic of the optical setup.

\textbf{Broadband transmission measurement}
The absorption spectrum (shown in Fig.~1a) were obtained from broadband transmission spectrum measured by a Bruker 80-V Fourier transform infrared spectrometer. A sample with thickness of $\sim 100 \mu$m and lateral size large than 3 mm was glued on a copper platform with a hollow hole. We measured the transmitted spectra of the sample and another hole with the same diameter for comparison. The transmission ratio $T$ was obtained by dividing the spectra of the sample and the hole.

\textbf{Wavelength-dependent THG} The dependence on THG emission wavelengths of $\chi^{(3)}$ shown in Fig.~1e was derived from THG measurements at varying excitation wavelengths. We used a Spectra-Physics TOPAS Prime optical parametric amplifier to generate excitation laser from 1200 nm to 2000 nm at 1 kHz repetition rate. The excitation laser was focused onto the MnPSe$_3$ 100 nm film sample using an optical lens with 125 mm focal length. The THG signals centered at emission wavelengths shorter (longer) than 550 nm were measured with the signal (idler) output of the optical parametric amplifier. The power of incident excitation laser was always kept at 0.4 mW.


\section*{Acknowledgements}
This work was supported by NSFC (No. 11888101), National Key Research and Development Program of China (NO. 2022YFA1403900, 2022YFA1403500, 2021YFA1400201), National Postdoctoral Program for Innovative Talents (No. BX20200016) and China Postdoctoral Science Foundation (No. 2020M680177, 2021M700257, 2021T140022, 2022M710232).


\clearpage

\end{document}